\documentclass{aa}
\usepackage{times}
\usepackage{epsfig}
\begin{document}
\thesaurus{04(12.07.1,12.04.1,08.02.3,08.12.2,08.16.2)}
\title{Ambiguities in fits of observed binary lens galactic microlensing events}
\author{M. Dominik\thanks{\emph{Present address:} Kapteyn Instituut,
Rijksuniversiteit Groningen, Postbus 800, NL-9700 AV Groningen,
The Netherlands 
(dominik@astro.rug.nl)}}
\institute{Institut f\"ur Physik, Universit\"at Dortmund, D-44221 Dortmund, Germany}
\date{Received ; accepted}
\maketitle
\begin{abstract}
For observed galactic microlensing events only one fit is usually presented, though, especially for
a binary lens, several fits may be possible. This has been shown for the MACHO LMC\#1 event
(Dominik \& Hirshfeld~\cite{DoHi2}). Here I discuss the strong binary lens events OGLE\#7 and DUO\#2.
It is shown that several models with a large variety of parameters are in
accordance with the photometric data. For most of the fits,
$1$-$\sigma$-bounds on the fit parameters are given. The variation of the parameters
within these bounds is in some cases considerable.
It is likely that other binary lens events which will occur will have properties similar to the 
discussed events. 

\end{abstract}

\section{Introduction}
Some of the observed galactic microlensing events show a signature of a binary
lens. The first such event discovered was MACHO LMC\#1,
where its binary nature has been discussed by Dominik \& Hirshfeld (\cite{DoHi1}, \cite{DoHi2}) and
by Rhie (\cite{Rhie}) and Rhie \& Bennett (\cite{RB}).
An explanation with a binary lens has also been proposed for OGLE\#6 by
Mao \& Di Stefano (\cite{MaoDis}). Neither of these events involves crossings of the
source trajectory with the caustics, which would result in sharp spikes\footnote{or at least a dramatic
rise and fall for larger sources}
in the light curves, which are a clear characteristic of a binary (or multiple)
lens.
For this reason, these events are called weak binary lens events
(according to Mao \& Paczy{\'n}ski (\cite{MP})).
In contrast, the events OGLE\#7 (Udalski et al.~\cite{OGLE7}), DUO\#2 
(Alard et al.~\cite{Alard}), and MACHO LMC\#9 (Bennett et al.~\cite{MACHOBin}) involve such
caustic crossings, and are therefore called strong binary lens events. 
The MACHO collaboration claims at least 5 additional binary lens events in their data
obtained with the alert system, namely 95-BLG-12, 96-BLG-3, 97-BLG-1, 97-BLG-28,
and 97-BLG-41 
(Stubbs et al.~\cite{MACHOAlert}).
Additional data for these events have been obtained by the MACHO collaboration 
for the OGLE\#7 event 
(Bennett et al.~\cite{MACHOOGLE2}, \cite{MACHOOGLE3}; Alcock et al.~\cite{MACHOOGLE1})
and by the PLANET collaboration
for the events 95-BLG-12, 97-BLG-28, and 97-BLG-41 
(Albrow et al.~1998a,b)\nocite{Planet1}\nocite{Planet2}.

It is a general feature that ambiguities may occur if the data has poor quality,
where poor quality means a bad sampling rate and an insufficient photometric accuracy.
However, except for the discussion of the MACHO LMC\#1 event (Dominik \& Hirshfeld 1996), only one
model has been presented for the binary lens microlensing events.
In this paper, it is shown that several binary lens models 
can fit the observed data  
for the events OGLE\#7 and DUO\#2 showing that these possible ambiguities exist among the observed events.
To draw the right conclusions from the observed data it is necessary
to find all models which fit the data. Therefore it is interesting to see which types of configurations produce which types of light curves. For the MACHO LMC\#1 event, 6 different configurations have been found which
can produce an asymmetric light curve of the desired form, so that whenever an event of this form occurs,
one should check for (at least) this 6 configurations. In fact, light curves of the type BL and of the type BA
can differ by less than 0.3\% which is much less than the photometric accuracy actually achieved by the
observing teams, so that this ambiguity remains
even with a perfect sampling of the data.
The arising ambiguities for binary lens events have also been addressed in a recent paper by Di~Stefano
\& Perna (\cite{DisPer}).

In this paper, configurations are shown which produce a double caustic crossing (OGLE\#7) and a double
caustic crossing with a following peak (DUO\#2). 
Though this paper deals with specific events, the types of light curves discussed here are expected to 
appear more frequently in the data. The configurations shown should be checked as possible solutions
for the fit whenever an event of  
this type occurs. Therefore, this analysis for specific events is instructive in general,
even if 
additional observations show that some of the models can be ruled out for these specific events.  
Better than what can be done by a simulation, the observed events show the actual quality of the data.

To extract the underlying physical quantities like the
masses of the lens objects,
it is of interest to see how well the observed data constrains the fit parameters. Except for MACHO LMC\#1
(Dominik \& Hirshfeld~\cite{DoHi1}, \cite{DoHi2}), no error bounds on the fit parameters have been given for binary lens
events. In this paper, these error bounds are given for most of the fits for OGLE\#7 and DUO\#2.

Section~2 reviews the basics of galactic microlensing with binary lenses, Sect.~3
gives the discussion of the OGLE\#7 event, while the DUO\#2 event is discussed
in Sect.~4. In Sect.~5 the results are summarized.

\section{Galactic microlensing with binary lenses}
\label{SSmodel}
Consider a lens at a distance $D_\mathrm{d}$ from the observer, a
source at a distance $D_\mathrm{s}$ from the observer, and let $D_\mathrm{ds}$
be the lens-source distance. The Einstein radius for a lens of
mass $M$ is then given by
\begin{equation}
r_\mathrm{E} = \sqrt{\frac{4GM}{c^2} \, \frac{D_\mathrm{d} D_\mathrm{ds}}{D_\mathrm{s}}}\,.
\end{equation}
Further consider an optical axis through observer and lens center and planes
perpendicular to it at the position of the lens (lens plane) and the source
(source plane) centered at the optical axis. 
Let the source be located at $\vec \eta = \frac{D_\mathrm{s}}{D_\mathrm{d}}\,r_\mathrm{E}\,\vec y$
in the source plane and let a light ray pass from
the source to the observer through the lens plane 
at $\vec \xi = r_\mathrm{E}\,\vec x$.  

For a binary lens consisting of a mass with fraction $m_1$ of the total mass
$M$ at $(r_\mathrm{hd},0)$ and a mass with fraction $m_2 = 1-m_1$ at 
$(-r_\mathrm{hd},0)$, where $r_\mathrm{hd} = \chi\,r_\mathrm{E}$, the coordinates
$\vec x$ and $\vec y$ are related by the lens equation
\begin{eqnarray}
y_1(x_1,x_2) & = &  x_1 - m_1\,\frac{x_1 -\chi}{(x_1-\chi)^2+x_2^2}
- \nonumber \\
& & - (1-m_1)\,\frac{x_1+\chi}{(x_1+\chi)^2+x_2^2}\,, \\
y_2(x_1,x_2) & = & x_2 - m_1\,\frac{x_2}{(x_1-\chi)^2+x_2^2}
- \nonumber \\
& & - (1-m_1)\,\frac{x_2}{(x_1+\chi)^2+x_2^2}\,,
\end{eqnarray}
which gives the true source position as a function of the observed image
position. For a given source position, there are either 3 or 5 images.
To obtain the images for a given source position $\vec y$ 
one has to solve the lens equation numerically.

The magnification $\mu$ of an image at $\vec x$ is given by the inverse of the
Jacobian determinant of the mapping, i.e.
\begin{equation}
\mu(\vec x) = \frac{1}{\left|\det \left(\frac{\partial \vec y}{\partial \vec x}\right)\right|}\,,
\label{mageq}
\end{equation} 
and the total magnification of a (point) source $\widetilde{\mu}(\vec y)$ 
is given as the sum of the magnifications of the individual images. 
A (point) source is located on a caustic if $\mu(\vec x)$ diverges for at least
one image $\vec x$.
The magnification of an extended source is obtained by integrating over the
point source magnifications. 

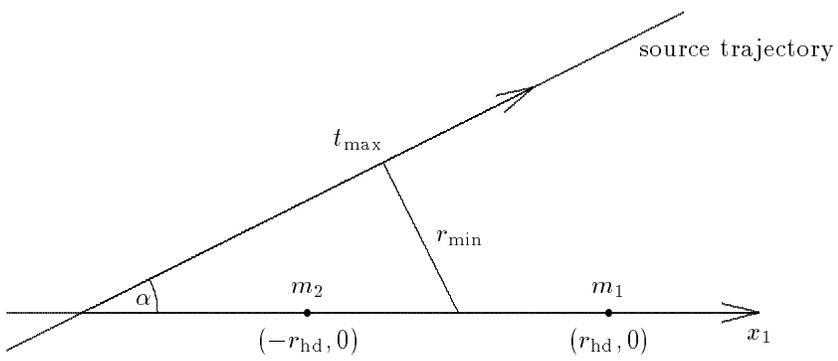
\begin{figure}[htb]
\vspace{5cm}
\caption{The geometry of a binary lens event}
\label{confgfigure}
\end{figure}

Let the source move on a straight line with a velocity $v_{\perp}$ projected onto the lens plane 
transverse to the line-of-sight 
between observer and lens, so that it moves 
one Einstein radius in the lens plane in the time
$t_\mathrm{E} = \frac{r_\mathrm{E}}{v_{\perp}}$.\footnote{Note that this
is equivalent to letting the lens move
with $v_{\perp}$ into the opposite direction. Further note that if the 
fixed source is on the right side of the lens trajectory, a corresponding
fixed lens is on the right side of the source trajectory.}
For the fits, a timescale $t_\mathrm{c} = 2\,t_\mathrm{E}$ is used.
In addition, the following parameters are used (see
Fig.~\ref{confgfigure})\footnote{These parameters coincide with those which have been
used in (Dominik \& Hirshfeld~\cite{DoHi1}, \cite{DoHi2})}:  
The source trajectory projected to the lens plane 
has the closest approach to the origin at the time $t_\mathrm{max}$
and the distance $r_\mathrm{min} = u_\mathrm{min}\,r_\mathrm{E}$.
The angle $\alpha$ is measured between the 
$x_1$-direction and the direction of motion of the source.
To get uniqueness in parameter space,  
the sign of the velocity of the
source is chosen so that the midpoint of the lens system 
is on the right hand side of
the line traced in time by the 
source and the 
parameter ranges are $u_{\min} \geq 0$,
$0 \leq \alpha < \pi$ and $0 \leq m_1 \leq 1$.  

The observed amplification of the source object as a function of time $A(t)$ has been
calculated efficiently 
by solving the lens equation for the whole source trajectory simultaneously
(Dominik~\cite{DoAstro}) and by determining the magnification from the images with
Eq.~(\ref{mageq}).

Another parametrization for binary lenses has been used by
Mao \& Di Stefano (\cite{MaoDis}). The main difference is that they refer
the closest approach of the source to the center of mass of the lens system
and not to the midpoint.
The translation between the parameter
sets is shown in Table~\ref{MaoDomtrans}.

\begin{table*}
\caption[ ]{The translation between different parameter sets for binary
lens models}
\begin{flushleft}
\begin{tabular}{ll}
\hline\noalign{\smallskip}
Mao \& Di Stefano $\to$ Dominik & Dominik $\to$ Mao \& Di Stefano \\
\noalign{\smallskip}\hline\noalign{\smallskip}
\rule[-1ex]{0ex}{3.5ex}
$\alpha [\mathrm{rad}] = \frac{\pi}{180} \vartheta [{}^{\circ}] $ &
$\vartheta [{}^{\circ}] =  \frac{180}{\pi} \alpha [\mathrm{rad}] $ \\
\rule[-1ex]{0ex}{3.5ex}
$m_1 = \frac{q}{q+1}$ &
$q = \frac{m_1}{1-m_1}$ \\
\rule[-1ex]{0ex}{3.5ex}
$\chi = {a / 2}$ &
$a = 2 \chi$ \\
\rule[-1ex]{0ex}{3.5ex}
$t_\mathrm{c} = 2 t_\mathrm{E}$ &
$t_\mathrm{E} = t_\mathrm{c}/2$ \\
\rule[-1ex]{0ex}{3.5ex}
$r_\mathrm{min} = b - \frac{a}{2}\, \frac{q-1}{q+1}\,\sin \vartheta$ &
$b = r_\mathrm{min} - \chi\,(1- 2 m_1)\,\sin \alpha$ \\
\rule[-1ex]{0ex}{3.5ex}
$t_\mathrm{max} = t_b - \frac{a}{2}\, \frac{q-1}{q+1}\,
t_\mathrm{E}\,
\cos \vartheta $ &
$t_b = t_\mathrm{max} - \chi\, (1-2 m_1)\,
\frac{t_\mathrm{c}}{2}\,\cos \alpha$ \\
\noalign{\smallskip}
\hline
\end{tabular}
\end{flushleft}
\label{MaoDomtrans}
\end{table*}

The amount of additional light contributed by other objects than the 'source' is described by the blending parameter
$f$. It gives the contribution of the light of the unlensed source to the total light.
If $A(t)$ denotes the amplification of the source, the observed amplification is given by
\begin{equation}
A_\mathrm{tot} = f A(t) + 1 - f\,.
\end{equation}

\section{An event with two caustic crossings: OGLE \#7}
\label{fits:OGLE7}

The OGLE\#7 event (Udalski et al.~\cite{OGLE7}, hereafter USM) is the first event observed which shows the 
signature of a strong binary lens. In contrast to the MACHO LMC\#1 event,
it is clear that a single lens cannot explain the data, because it
will produce
neither spikes nor a plateau formed like a U, irrespective of whether the
source is a binary or not or whether the objects are extended.
A first fit to the data has been presented in~USM. 
However, several binary lens fits  for these data are possible.
In Tables~\ref{OGLE7fits1} and~\ref{OGLE7fits2}, I present the
results of 8 different fits. The $\chi^2$ corresponds to the 
original errors given by the OGLE collaboration. The errors on the
parameters
correspond to projections of the hypersurface $\Delta \chi^2 
= \chi^2 - \chi^2_\mathrm{min} = 1$ onto the axes in parameter space.
The characteristic time ranges from $162~\mbox{d}$ (BL0) 
to $1470~\mbox{d}$ (BL6) which means that the expected total mass
($\propto t_\mathrm{c}^2$)
varies by a factor of 80. The mass ratio ranges from about 1
(BL0, BL6, BL7) to 16 (BL). The contribution of external light (blending)
at the mimimum light ranges from 17~\% (BL9) to 94~\% (BL).
Note that some of the errors on $t_\mathrm{c}$ are large: For
BL2 the 1-$\sigma$ error is about 25~\% of the value, for
BL7, BL1, BL3, and BL9 the 1-$\sigma$ errors are also about
10--20~\% of the value of $t_\mathrm{c}$. The second pair of caustic crossings
for the BL-fit and the caustic crossings on the triangle shaped caustic
for the BL6-fit posed additional
problems for the calculation of the
error bounds, so that no error bounds are shown for these fits.

The fit of USM corresponds to the BL0-fit. Table~\ref{o7ogle}
shows the parameters of the BL0-fit in the parametrization used in
USM and the parameters quoted there. Note that their $m_0$
is the magnitude of the lensed component, while $m_\mathrm{base}$ 
corresponds to the minimum light observed, i.e. the light from the lensed
and the unlensed component. These quantities are therefore related by
\begin{equation}
m_0 = - m_\mathrm{base} - 2.5\,\mathrm{lg}\,f\,.
\end{equation}

If one accepts the given error bars of the data points, one sees
that the values of $\chi^2_\mathrm{min}$ are very bad (see 
Table~\ref{OGLE7probU}). If one adapts the size of the error bars to the 
tail data points in the same way as for the MACHO LMC\#1 event (Dominik \& Hirshfeld~\cite{DoHi2}),
but with Gaussian errors\footnote{As shown for the MACHO LMC\#1 event (Dominik~\cite{DoDiss})
the results do not differ much if one uses a larger rescaling factor
with a distribution with smaller tail or a smaller rescaling factor
with a distribution with larger tail. Since there are only a few
data points in the tail region (32), the extraction of non-Gaussian behaviour
is not successful (Dominik~\cite{DoDiss}).}, defining 
all data points for $t < 900~\mathrm{d}$ as belonging to the 'tail', one obtains a
scaling factor of $\gamma = 1.597$, which
means that the errors are enlarged by about 60~\%.
The probabilities for the fits become $\geq 84~\%$ for all fits
and $\geq 95~\%$ for all fits except for BL3 and BL9.
If one looks carefully at the light curve and the data, one
sees that the data point in the tail at $t = 806.60603$ and
$I = 17.794$ contributes about 33 to the
total $\chi^2$. If one omits this discrepant point, one gets a
rescaling factor of $\gamma = 1.247$ and the
results shown in Table~\ref{OGLE7probNpt}.
If one assumes that the discrepant point in the tail is due to a
measurement error, and taking into account that the error bars may be about
15-25~\% too small, one can accept all 8 models (see Table~\ref{OGLE7probNpt}).
If one does not ascribe the discrepant point to  a measurement error,
one has to accept the large rescaling factor to allow the tail to be 
constant. Also in this case, all the fits are acceptable.
Note that the errors may show some non-Gaussian behaviour. This behaviour
is effectively absorbed into the rescaling factor in my analysis.

\begin{table*}
\caption{OGLE \#7: Fits for a strong binary lens event}
\label{OGLE7fits1}
\begin{flushleft}
\begin{tabular}{lcccc}
\hline\noalign{\smallskip}
parameter & BL2 & BL0 & BL1 & BL7 \\
\noalign{\smallskip}\hline\noalign{\smallskip}
\rule[-1ex]{0ex}{3.5ex}$t_\mathrm{c}$ [d]& 
$345_{-82}^{+95}$ & 
$161.8_{-6.9}^{+6.8}$ & 
$291_{-19}^{+41}$ &
$350_{-39}^{+52}$ \\ 
\rule[-1ex]{0ex}{3.5ex}$t_\mathrm{max}$ [d]&
$1147.6_{-7.4}^{+4.7}$ & 
$1173.3_{-2.0}^{+2.5}$ &
$1164.6_{-1.5}^{+1.2}$ &
$1150.86_{-10.5}^{+0.47}$ \\ 
\rule[-1ex]{0ex}{3.5ex}$\chi$ & 
$0.400_{-0.028}^{+0.041}$ &
$0.565_{-0.010}^{+0.011}$ &
$0.387_{-0.013}^{+0.030}$ &
$0.496_{-0.011}^{+0.021}$ \\ 
\rule[-1ex]{0ex}{3.5ex}$m_1$ &
$0.268_{-0.054}^{+0.082}$ &
$0.506_{-0.014}^{+0.029}$ &
$0.397_{-0.015}^{+0.089}$ &
$0.518_{-0.043}^{+0.022}$ \\ 
\rule[-1ex]{0ex}{3.5ex}$\alpha$ [rad]&
$2.517_{-0.062}^{+0.028}$ &
$2.297_{-0.048}^{+0.044}$ &
$1.635_{-0.026}^{+0.025}$ &
$0.135_{-0.099}^{+0.025}$ \\ 
\rule[-1ex]{0ex}{3.5ex}$u_\mathrm{min}$ & 
$0.013_{-0.012}^{+0.006}$ &
$0.048_{-0.006}^{+0.013}$ &
$0.01100_{-0.020}^{+0.00088}$ &
$0.539_{-0.060}^{+0.054}$ \\ 
\rule[-1ex]{0ex}{3.5ex}$m_\mathrm{base}$ &
$-17.5299_{-0.0077}^{+0.0079}$ &
$-17.5171_{-0.0070}^{+0.0071}$ &
$-17.5260_{-0.0088}^{+0.0077}$ &
$-17.5487_{-0.013}^{+0.0091}$ \\ 
\rule[-1ex]{0ex}{3.5ex}$f$ &
$0.241_{-0.041}^{+0.066}$ &
$0.557_{-0.022}^{+0.020}$ &
$0.243_{-0.014}^{+0.017}$ &
$0.618_{-0.019}^{+0.019}$ \\ 
\noalign{\smallskip}\hline\noalign{\smallskip}
\rule[-1ex]{0ex}{3.5ex}$\chi^2_\mathrm{min}$ & 
133.94 & 
137.73 &
137.93 &
140.25 \\ \noalign{\smallskip}\hline
\end{tabular}
\end{flushleft}
\end{table*}

\begin{table*}
\caption{OGLE \#7: Fits for a strong binary lens event}
\label{OGLE7fits2}
\begin{flushleft}
\begin{tabular}{lcccc}
\hline\noalign{\smallskip}
parameter & BL6 & BL & BL3 & BL9 \\
\noalign{\smallskip}\hline\noalign{\smallskip}
\rule[-1ex]{0ex}{3.5ex}$t_\mathrm{c}$ [d]& 
$1470.84_{}^{}$ & 
$1299.68_{}^{}$ & 
$440.56_{-26}^{+43}$ &
$495_{-70}^{+73}$ \\ 
\rule[-1ex]{0ex}{3.5ex}$t_\mathrm{max}$ [d]&
$2149.92_{}^{}$ & 
$1380.10_{}^{}$ &
$1168.4_{-2.7}^{+2.1}$ &
$1160.93_{-2.4}^{+0.95}$ \\ 
\rule[-1ex]{0ex}{3.5ex}$\chi$ & 
$0.2813_{}^{}$ &
$0.3725_{}^{}$ &
$0.854_{-0.039}^{+0.044}$ &
$0.736_{-0.010}^{+0.014}$ \\
\rule[-1ex]{0ex}{3.5ex}$m_1$ &
$0.510_{}^{}$ &
$0.059_{}^{}$ &
$0.759_{-0.11}^{+0.081}$ &
$0.938_{-0.009}^{+0.024}$ \\
\rule[-1ex]{0ex}{3.5ex}$\alpha$ [rad]&
$2.050_{}^{}$ &
$0.414_{}^{}$ &
$1.567_{-0.025}^{+0.031}$ &
$1.942_{-0.054}^{+0.043}$ \\
\rule[-1ex]{0ex}{3.5ex}$u_\mathrm{min}$ & 
$0.649_{}^{}$ &
$0.068_{}^{}$ &
$0.316_{-0.064}^{+0.077}$ &
$0.086_{-0.023}^{+0.027}$ \\
\rule[-1ex]{0ex}{3.5ex}$m_\mathrm{base}$ &
$-17.5868_{}^{}$ &
$-17.5555_{}^{}$ &
$-17.5538_{-0.0096}^{+0.0079}$ &
$-17.639_{-0.030}^{+0.031}$ \\
\rule[-1ex]{0ex}{3.5ex}$f$ &
$0.667_{}^{}$ &
$0.057_{}^{}$ &
$0.578_{-0.064}^{+0.052}$ &
$0.829_{-0.043}^{+0.032}$ \\ 
\noalign{\smallskip}\hline\noalign{\smallskip}
\rule[-1ex]{0ex}{3.5ex}$\chi^2_\mathrm{min}$ & 
141.49 & 
143.41 &
150.95 &
160.06 \\ \noalign{\smallskip}\hline
\end{tabular}
\end{flushleft}
\end{table*}

\begin{table}
\caption{OGLE \#7: The BL0-fit and the fit of USM}
\label{o7ogle}
\begin{flushleft}
\begin{tabular}{lcc}
\hline\noalign{\smallskip}
parameter & BL0 & Udalski et al. \\
\noalign{\smallskip}\hline\noalign{\smallskip}
\rule[-1ex]{0ex}{3.5ex}$t_\mathrm{E}$ [d]& 
$80.88$ & 
$80$ \\ 
\rule[-1ex]{0ex}{3.5ex}$t_b$ [d]&
$1172.89$ & 
$1172.5$ \\
\rule[-1ex]{0ex}{3.5ex}$a$ & 
$1.1309$ &
$1.14$ \\ 
\rule[-1ex]{0ex}{3.5ex}$q$ &
$1.024$ &
$1.02$ \\ 
\rule[-1ex]{0ex}{3.5ex}$\theta$ $[{}^{\circ}]$&
$131.68$ &
$138.3$ \\
\rule[-1ex]{0ex}{3.5ex}$b$ & 
$0.0532$ &
$0.050$ \\
\rule[-1ex]{0ex}{3.5ex}$m_0$ &
$18.152$ &
$18.1$ \\ 
\rule[-1ex]{0ex}{3.5ex}$f$ &
$0.5573$ &
$0.56$ \\ \noalign{\smallskip}\hline
\end{tabular}
\end{flushleft}
\end{table} 

\begin{table*}
\caption{OGLE \#7: $\chi^2_\mathrm{min}$ and the
corresponding probability without rescaling}
\label{OGLE7probU}
\begin{flushleft}
\begin{tabular}{lcccccccc}
\hline\noalign{\smallskip}
 & BL2 & BL0 & BL1 & BL7 & BL6 & BL & BL3 & BL9 \\
\noalign{\smallskip}\hline\noalign{\smallskip}
\rule[-1ex]{0ex}{3.5ex}$\chi^2_\mathrm{min}$& 
133.94 & 137.73 & 137.93 & 140.25 & 141.49 & 143.41 & 150.95 & 160.06 
\\ 
\rule[-1ex]{0ex}{3.5ex}$\sqrt{2\chi_\mathrm{min}^2}-\sqrt{2n-1}$&
4.160 & 4.390 & 4.402 & 4.542 & 4.615 & 4.729 & 5.169 & 5.685 
\\ 
\rule[-1ex]{0ex}{3.5ex}$P(\chi^2 \geq \chi^2_\mathrm{min})$ & 
$2\cdot 10^{-5}$ & $6\cdot 10^{-6}$ & $5\cdot 10^{-6}$ & $3\cdot 10^{-6}$ &
$2\cdot 10^{-6}$ & $1\cdot 10^{-6}$ & $1\cdot 10^{-7}$ & $7\cdot 10^{-9}$
\\ \noalign{\smallskip}\hline
\end{tabular}
\end{flushleft}
\end{table*}

\begin{table*}
\caption{OGLE \#7: $\chi^2_\mathrm{min}$ and the
corresponding probability without discrepant point 
with rescaling}
\label{OGLE7probNpt}
\begin{flushleft}
\begin{tabular}{lcccccccc}
\hline\noalign{\smallskip}
 & BL2 & BL0 & BL1 & BL7 & BL6 & BL & BL3 & BL9 \\
\noalign{\smallskip}\hline\noalign{\smallskip}
\rule[-1ex]{0ex}{3.5ex}$\chi^2_\mathrm{min}$& 
64.98 & 66.92 & 67.55 & 68.65 & 69.71 & 70.37 & 76.60 & 80.86 
\\
\rule[-1ex]{0ex}{3.5ex}$\sqrt{2\chi_\mathrm{min}^2}-\sqrt{2n-1}$&
-0.807 & -0.638 & -0.583 & -0.489 & -0.399 & -0.343 & 0.171 & 0.510 
\\
\rule[-1ex]{0ex}{3.5ex}$P(\chi^2 \geq \chi^2_\mathrm{min})$ & 
79~\% & 74~\% & 72~\% & 69~\% & 66~\% & 63~\% & 43~\% & 30~\%
\\ \noalign{\smallskip}\hline
\end{tabular}
\end{flushleft}
\end{table*}

The light curves for the fits are shown in Fig.~\ref{o7lcblc2}
and
the structure of the caustics together with the
source trajectory can be seen in Fig.~\ref{o7cfblc2}.
The distance scale used is projected 
Einstein radii 
($r_\mathrm{E}' = \frac{D_\mathrm{s}}{D_\mathrm{d}}\,r_\mathrm{E}$)
of the total mass in the source plane. The intersections
of the source trajectory with the caustics and the projected positions of the
lens objects are indicated by small crosses. The tip of the arrow on the
trajectory denotes the closest approach to the coordinate origin.

\begin{figure*}[htb]
\vspace{20cm}
\caption{OGLE\#7: Light curves for the different binary lens fits 
and measured data points in I-band. From left to right, top to bottom:
(a) BL2, (b) BL0, (c) BL1, (d) BL7, (e) BL6, (f) BL, (g) BL3, (h) BL9.}
\label{o7lcblc2}
\end{figure*}

\begin{figure*}[htb]
\vspace{20.5cm}
\caption{OGLE\#7. Caustics and source trajectory for the different
binary lens fits.
The distance is measured in projected Einstein radii of the total mass. 
Small crosses indicate crossings of the source trajectory with the caustics
and the projected positions of the point masses. From left to right, top to bottom:
(a) BL2, (b) BL0, (c) BL1, (d) BL7, (e) BL6, (f) BL, (g) BL3, (h) BL9.} 
\label{o7cfblc2}
\end{figure*}

Additional data for this event has been obtained by the MACHO collaboration 
(Bennett et al.~\cite{MACHOOGLE2}, \cite{MACHOOGLE3}; Alcock et al.~\cite{MACHOOGLE1}).
The main feature of this data is a data point (in two bands) on the fall of the second caustic crossing,
which can only be fitted by including a finite source size. 
If one omits this point,
models with configurations similar to BL7, BL2, BL1, BL3, and BL0 give good fits (in descending order),\footnote{
The MACHO collaboration did not make this data available in electronic form, these statements are
based on data obtained by reading off the values from their figures.}  BL is 
marginal and BL6 and BL9 are excluded. If one includes the point on the fall of the second caustic
crossing, the 
approximation of a point source fails. If one allows for a finite source size, there are still several
possible models remaining, with parameters similar to the above point-source fits: 
Models similar to BL1, BL0, BL2, and BL3 give successful fits (in descending order of goodness-of-fit).
The additional MACHO data is therefore not sufficient to break the ambiguities for this event completely.
Nevertheless, it is advisory to test all of the above types of configurations for an event like OGLE\#7
to see whether there are fit ambiguities and to find a suitable model.

\section{An event with two caustic crossings and an additional peak: DUO \#2}
\label{fits:DUO2}

The DUO\#2 event has been reported by Alard et al. (\cite{Alard}),
hereafter AMG, where a fit with
a strong binary lens is presented. The corresponding fit parameters 
using my parameter set are shown
in Table~\ref{DUO2orig}. For 116~data points, a $\chi^2$ of 89 is
reached, where the source is treated as extended ($R_\mathrm{src} \la
10^{-2}$) and limb-darkened with $\widetilde{u}_\mathrm{blue} = 0.6$ and
$\widetilde{u}_\mathrm{red} = 0.5$, where the brightness profile of the
source is of the form
\begin{equation}
f(r,\widetilde{u}) = \widetilde{u}\sqrt{1-r^2}+1-\widetilde{u}\,,
\end{equation}
$r$ being the ratio of the actual radius and the total radius.

Here I show additional possible fits using
a static binary lens and a point source. 
I have omitted one occurence of a data point which
appeared twice in the data I have received from C. Alard, 
so that I use 115 data points and not 116~data
points as in AMG. Moreover, I use the magnitude values for the
fit and not the amplification values as used for the fit in AMG
(S. Mao, private communication). From this, only a small difference
results (Dominik~\cite{DoDiss}), the $\chi^2_\mathrm{min}$
differs only by about $2$ units and the fit parameters show only small
differences which are well within the bounds corresponding to
$\Delta \chi^2 = 1$. 
The analysis of the tail data points ($t < 70~\mathrm{d}$ or $t > 100~\mathrm{d}$)
yielded most-likely scaling factors $\gamma_\mathrm{blue} = 0.912$ for 63
data points and $\gamma_\mathrm{red} = 0.828$ for 25 data points. These
values are close to 1 (note that there are only a few tail data points),
so that the 
tail seems to be consistent with a constant brightness and no further
scaling is used in this discussion.
The results of the fits are shown in 
Tables~\ref{DUO2fits1} and~\ref{DUO2fits2}. The quoted error bounds
correspond to projections of the hypersurface $\Delta \chi^2 = 1$ onto
the axes in parameter space. An asterisk (${}^{\ast}$) denotes that
the numerical routines have ended up at a jump discontinuity. It is not
yet clear if this is a real effect or if it is due to difficulties in the
computation.
There appear additional minima, whose $\Delta \chi^2 \leq 1$-regions
include other minima with smaller $\chi^2$. In particular, this is a
problem for the BL1-, BL2-, and the BL4-fit which also makes the
calculation of the error bounds difficult, so that they are not
shown in the tables for these fits. This behavior 
is influenced by the
fact that there are only a few data points to constrain the shape of the
light curve in the peak region.

One sees that the BL- and the BL4-fit give good explanations of the observed
data, while the BL2-, BL3-, and the BL5-fit give worse results, although they
are not totally excluded. The BL1-fit gives such a low probability that
it is excluded. Light curves of the peak region ($70~\mbox{d} \leq t \leq
100~\mbox{d}$) for both spectral bands together with the data points
are shown in Figs.~\ref{d2lcbl} and~\ref{d2lcbl2}, where the upper curve
refers to the blue band and the lower curve refers to the red band. Note
that the largest differences between the BL-fit and the extended source fit
near the BL-parameters occur in the peak after the caustic crossings, not
in the caustic crossings themselves.

\begin{figure*}[htb]
\vspace{17cm}
\caption{DUO \#2: Light curves for the different binary lens fits
 and measured data points.
Light curve for the blue spectral band on the top and 
light curve for the red spectral band on the bottom.
From left to right, top to bottom: (a) BL, (b) BL4, (c) BL2, (d) BL3.
}
\label{d2lcbl}
\end{figure*}

\begin{figure*}[htb]
\vspace{8cm}
\caption{DUO \#2: Light curves for the different binary lens fits
 and measured data points.
Light curve for the blue spectral band on the top and 
light curve for the red spectral band on the bottom.
From left to right: (a) BL5, (b) BL1.
}
\label{d2lcbl2}
\end{figure*}

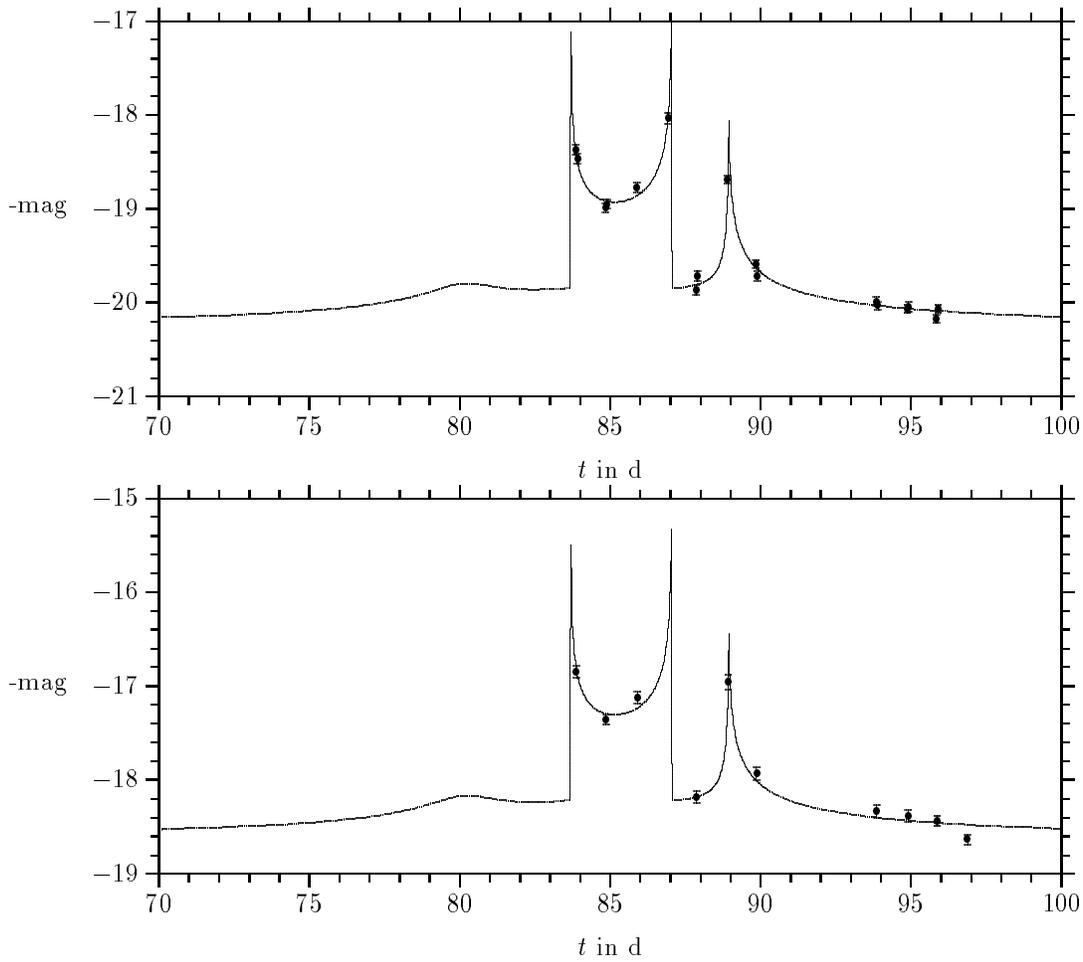
\begin{figure}[htb]
\vspace{8cm}
\caption{DUO \#2: Light curves for the fit near BL with an
extended source and the measured data points.
Light curve for the blue spectral band on the top and 
light curve for the red spectral band on the bottom.}
\label{d2lcblext}
\end{figure}

The configurations for the different models are shown in Fig.~\ref{d2cfbl},
where the caustics are shown together with the trajectory
of the moving point source. The projected positions of the lens objects and
the intersections of the source trajectory with the caustics are indicated
by small crosses. All distances are measured in Einstein radii $r_\mathrm{E}$
if the projection to the lens plane is considered or in projected
Einstein radii $r_\mathrm{E}'$ if the projection to the source plane is considered.

\begin{figure*}[htb]
\vspace{20cm}
\caption{DUO \#2: Caustics and source trajectory for the different binary
lens fits. From left to right, top to bottom:
(a) BL, (b) BL4, (c) BL2, (d) BL3, (e) BL5, (f) BL1.}
\label{d2cfbl}
\end{figure*}

\begin{table}
\caption{DUO \#2: Fit of AMG}
\label{DUO2orig}
\begin{flushleft}
\begin{tabular}{lc}
\hline\noalign{\smallskip}
parameter & DUO\#2 in AMG \\
\noalign{\smallskip}\hline\noalign{\smallskip}
\rule[-1ex]{0ex}{3.5ex}$t_\mathrm{c}$ [d]& 
17 \\ 
\rule[-1ex]{0ex}{3.5ex}$t_\mathrm{max}$ [d]&
85.19 \\
\rule[-1ex]{0ex}{3.5ex}$\chi$ & 
0.605 \\
\rule[-1ex]{0ex}{3.5ex}$m_1$ &
0.752  \\
\rule[-1ex]{0ex}{3.5ex}$\alpha$ [rad]&
1.490 \\ 
\rule[-1ex]{0ex}{3.5ex}$u_\mathrm{min}$ & 
0.096 \\ 
\rule[-1ex]{0ex}{3.5ex}$m_\mathrm{base,blue}$ &
-20.21 \\
\rule[-1ex]{0ex}{3.5ex}$m_\mathrm{base,red}$ &
-18.55 \\
\rule[-1ex]{0ex}{3.5ex}$f_\mathrm{blue}$ &
0.73 \\ 
\rule[-1ex]{0ex}{3.5ex}$f_\mathrm{red}$ &
0.70 \\ \noalign{\smallskip}\hline
\end{tabular}
\end{flushleft}
\end{table}

\begin{table*}
\caption{DUO \#2: Fits for a strong binary lens event}
\label{DUO2fits1}
\begin{flushleft}
\begin{tabular}{lccc}
\hline\noalign{\smallskip}
parameter & BL & BL4 & BL2 \\
\noalign{\smallskip}\hline\noalign{\smallskip}
\rule[-1ex]{0ex}{3.5ex}$t_\mathrm{c}$ [d]& 
$18.9_{-1.5}^{+2.5}$ & 
$23.79_{}^{}$ &
$16.40_{}^{}$ \\ 
\rule[-1ex]{0ex}{3.5ex}$t_\mathrm{max}$ [d]&
$85.234_{-0.062}^{+0.075}$ & 
$89.247_{}^{}$ &
$88.351_{}^{}$ \\
\rule[-1ex]{0ex}{3.5ex}$\chi$ & 
$0.6434_{-0.0011^{\ast}}^{+0.050}$ &
$0.4058_{}^{}$ &
$0.4571_{}^{}$ \\
\rule[-1ex]{0ex}{3.5ex}$m_1$ &
$0.759_{-0.026}^{+0.005}$ &
$0.176_{}^{}$ &
$0.829_{}^{}$ \\ 
\rule[-1ex]{0ex}{3.5ex}$\alpha$ [rad]&
$1.496_{-0.024}^{+0.023}$ &
$0.670_{}^{}$ &
$2.576_{}^{}$ 
 \\ 
\rule[-1ex]{0ex}{3.5ex}$u_\mathrm{min}$ & 
$0.157_{-0.043^{\ast}}^{+0.025^{\ast}}$ &
$0.00684_{}^{}$ &
$0.0406_{}^{}$ \\ 
\rule[-1ex]{0ex}{3.5ex}$m_\mathrm{base,blue}$ &
$-20.2077_{-0.0059}^{+0.0061}$ &
$-20.2087_{}^{}$ &
$-20.2046_{}^{}$ \\
\rule[-1ex]{0ex}{3.5ex}$m_\mathrm{base,red}$ &
$-18.575_{-0.011}^{+0.011}$ &
$-18.5777_{}^{}$ &
$-18.5733_{}^{}$ \\
\rule[-1ex]{0ex}{3.5ex}$f_\mathrm{blue}$ &
$0.708_{-0.048}^{+0.035}$ &
$0.260_{}^{}$ &
$0.418_{}^{}$ \\ 
\rule[-1ex]{0ex}{3.5ex}$f_\mathrm{red}$ &
$0.702_{-0.054}^{+0.044}$ &
$0.266_{}^{}$ &
$0.441_{}^{}$ \\ 
\noalign{\smallskip}\hline\noalign{\smallskip}
\rule[-1ex]{0ex}{3.5ex}$\chi^2_\mathrm{min}$ & 
109.45 & 
111.57 &
131.17 \\
\rule[-1ex]{0ex}{3.5ex}$n$ = \# d.o.f. & 
105 & 
105 &
105 \\ 
\rule[-1ex]{0ex}{3.5ex}$\sqrt{2\chi^2_\mathrm{min}}-\sqrt{2n-1}$ & 
0.338 & 
0.481 &
1.740 \\ 
\rule[-1ex]{0ex}{3.5ex}$P(\chi^2 \geq \chi^2_\mathrm{min})$ & 
37~\% & 
32~\% &
4~\% \\ \noalign{\smallskip}\hline
\end{tabular}
\end{flushleft}
\end{table*} 

\begin{table*}
\caption{DUO \#2: Fits for a strong binary lens event}
\label{DUO2fits2}
\begin{flushleft}
\begin{tabular}{lccc}
\hline\noalign{\smallskip}
parameter & BL3 & BL5 & BL1 \\
\noalign{\smallskip}\hline\noalign{\smallskip}
\rule[-1ex]{0ex}{3.5ex}$t_\mathrm{c}$ [d]& 
$14.95_{-0.93}^{+0.94}$ &
$12.30_{-0.57}^{+0.21^{\ast}}$ & 
$14.40_{}^{}$ 
 \\ 
\rule[-1ex]{0ex}{3.5ex}$t_\mathrm{max}$ [d]&
$85.69_{-0.24}^{+0.15}$ &
$84.29_{-0.28}^{+0.06^{\ast}}$ & 
$85.99_{}^{}$ 
 \\ 
\rule[-1ex]{0ex}{3.5ex}$\chi$ & 
$0.7832_{-0.0078}^{+0.0068}$ &
$0.495_{-0.005^{\ast}}^{+0.009}$ &
$0.649_{}^{}$ 
 \\ 
\rule[-1ex]{0ex}{3.5ex}$m_1$ &
$0.263_{-0.024}^{+0.036}$ &
$0.195_{-0.031}^{+0.030}$ &
$0.394_{}^{}$ 
 \\ 
\rule[-1ex]{0ex}{3.5ex}$\alpha$ [rad]&
$0.536_{-0.030}^{+0.022}$ &
$2.260_{-0.044^{\ast}}^{+0.071}$ &
$0.423_{}^{}$ 
 \\ 
\rule[-1ex]{0ex}{3.5ex}$u_\mathrm{min}$ & 
$0.053_{-0.017}^{+0.011}$ &
$0.161_{-0.016}^{+0.017}$ &
$0.261_{}^{}$ 
 \\ 
\rule[-1ex]{0ex}{3.5ex}$m_\mathrm{base,blue}$ &
$-20.2075_{-0.0062}^{+0.0065}$ &
$-20.1980_{-0.0057}^{+0.0057}$ &
$-20.2059_{}^{}$ 
 \\ 
\rule[-1ex]{0ex}{3.5ex}$m_\mathrm{base,red}$ &
$-18.574_{-0.011}^{+0.011}$ &
$-18.564_{-0.011}^{+0.011}$ &
$-18.573_{}^{}$ 
 \\ 
\rule[-1ex]{0ex}{3.5ex}$f_\mathrm{blue}$ &
$0.886_{-0.037^{\ast}}^{+0.024}$ &
$0.548_{-0.042}^{+0.042}$ &
$0.803_{}^{}$ 
 \\ 
\rule[-1ex]{0ex}{3.5ex}$f_\mathrm{red}$ &
$0.872_{-0.042}^{+0.037}$ &
$0.529_{-0.044}^{+0.044}$ &
$0.795_{}^{}$ 
 \\  
\noalign{\smallskip}\hline\noalign{\smallskip}
\rule[-1ex]{0ex}{3.5ex}$\chi^2_\mathrm{min}$ & 
134.96 & 
137.63 &
168.39 
 \\ 
\rule[-1ex]{0ex}{3.5ex}$n$ = \# d.o.f. & 
105 & 
105 &
105 \\ 
\rule[-1ex]{0ex}{3.5ex}$\sqrt{2\chi^2_\mathrm{min}}-\sqrt{2n-1}$ & 
1.973 & 
2.134 &
3.89 \\ 
\rule[-1ex]{0ex}{3.5ex}$P(\chi^2 \geq \chi^2_\mathrm{min})$ & 
2~\% & 
2~\% &
$5\cdot 10^{-5}$ \\ \noalign{\smallskip}\hline
\end{tabular}
\end{flushleft}
\end{table*}

Taking into account the error bounds, the BL-fit seems to coincide with
the fit in AMG, except for the small lower bound on $\chi$, which
however may be relict of computational problems. Near the parameters
for the BL-fit, I have started a fit including an extended source.
As for the fit of AMG, the 
source brightness profile has been fixed to a limb-darkening 
profile with $\widetilde{u}_\mathrm{blue} = 0.6$ and 
$\widetilde{u}_\mathrm{red} = 0.5$.
The resulting fit parameters are shown in Table~\ref{DUO2extsrc}.
Note that this fit gives only a slightly better $\chi^2_\mathrm{min}$ than
the BL-fit with a point source. The parameters coincide with the BL-fit
as well as with the fit of AMG.
It it not clear to me how the $\chi^2_\mathrm{min} = 89$ of AMG
is reached. For their fit parameters (S. Mao, private communication),
I obtain a $\chi^2$ which is larger than the $\chi^2_\mathrm{min}$ for the
extended source fit.

\begin{table}
\caption{DUO \#2: Fit for a strong binary lens and an extended source}
\label{DUO2extsrc}
\begin{flushleft}
\begin{tabular}{lc}
\hline\noalign{\smallskip}
parameter & binary lens, extended source \\
\noalign{\smallskip}\hline\noalign{\smallskip}
\rule[-1ex]{0ex}{3.5ex}$t_\mathrm{c}$ [d]& 
19.1 \\ 
\rule[-1ex]{0ex}{3.5ex}$t_\mathrm{max}$ [d]&
85.18 
 \\ 
\rule[-1ex]{0ex}{3.5ex}$\chi$ & 
0.626 
 \\ 
\rule[-1ex]{0ex}{3.5ex}$m_1$ &
0.738
 \\ 
\rule[-1ex]{0ex}{3.5ex}$\alpha$ [rad]&
1.447 
 \\ 
\rule[-1ex]{0ex}{3.5ex}$u_\mathrm{min}$ & 
0.159 
 \\ 
\rule[-1ex]{0ex}{3.5ex}$m_\mathrm{base,blue}$ &
-20.207 
 \\ 
\rule[-1ex]{0ex}{3.5ex}$m_\mathrm{base,red}$ &
-18.575 
 \\ 
\rule[-1ex]{0ex}{3.5ex}$f_\mathrm{blue}$ &
0.665
 \\ 
\rule[-1ex]{0ex}{3.5ex}$f_\mathrm{red}$ &
0.657 
 \\  
\rule[-1ex]{0ex}{3.5ex}$R_\mathrm{src}$ &
0.0012
 \\  
\noalign{\smallskip}\hline\noalign{\smallskip}
\rule[-1ex]{0ex}{3.5ex}$\chi^2_\mathrm{min}$ & 
107.28 
 \\ 
\rule[-1ex]{0ex}{3.5ex}$n$ = \# d.o.f. & 
104 \\ 
\rule[-1ex]{0ex}{3.5ex}$\sqrt{2\chi^2_\mathrm{min}}-\sqrt{2n-1}$ & 
0.260 \\
\rule[-1ex]{0ex}{3.5ex}$P(\chi^2 \geq \chi^2_\mathrm{min})$ & 
40~\% \\ \noalign{\smallskip}\hline
\end{tabular}
\end{flushleft}
\end{table}

A peak near a caustic
crossing can also be modeled in a different way with a rotating binary
lens. With parameters similar to those used
 for the model BL0 for OGLE\#7, a peak after the caustic crossing can be
modeled by including the rotation (Dominik~\cite{DoRob}). 

In AMG, it has been adressed that the observation of the position of the blend together with the observation
of the shift in the centroid of light gives an additional constraint. In fact, with these observations, a
constraint on the blending parameter $f$ is obtained. The observed data for the centroid shift and the position
of the blend should rule out the models BL4 and BL2, leaving the model BL (which had been proposed
by AMG) as the most likely interpretation, though the BL5 model does not seem to be completely excluded.

\section{Summary of results}
It has been shown that there are several models which fit the observed 
photometric data for the strong
binary lens events OGLE\#7 and DUO\#2, as well as for the weak binary lens
event MACHO LMC\#1 (Dominik \& Hirshfeld~\cite{DoHi2}). A large variety of timescales results,
so that the expectation value for the mass (see Dominik~\cite{DoMass}) differs by a
factor of 80 for the different fits for OGLE\#7. In addition, the uncertainty
in $t_\mathrm{c}$ for a given fit as given by the 1-$\sigma$-bound is as large as
25~\% for OGLE\#7, which is substantially larger than for MACHO LMC\#1 or
events described by single point-mass lenses.

While ambiguities for DUO\#2 are due to the bad sampling rate, the light curves for the different
models are much more similar for OGLE\#7 and MACHO LMC\#1, so that even for a good (or perfect) sampling,
ambiguities may occur due to the limited photometric precision.
Whether there are ambiguities or not for a given photometric precision depends on the distinctive
features that are present in the light curve. The asymmetric peak is a less distinctive feature than
the double caustic crossing of OGLE\#7, an asymmetric peak can also be modeled by a binary source, while the
spikes at the caustic crossing clearly indicate a lens binary (or multiple). The additional peak for DUO\#2 is 
an additional distinctive feature compared to OGLE\#7.
More distinctive features can be present if one can observe the finite source size (Alcock et al.~\cite{MACHOExt};
Albrow et al.~\cite{Planet2}), or rotation effects (Dominik~\cite{DoRob}) in the photometric
data, where rotation effects can be the earth-sun
parallax (Alcock et al.~\cite{MACHOExt}) 
the rotation of a binary source (Paczy{\'n}ski~\cite{PacRot}) or of a
binary lens (example yet to be detected). 
Some of the ambiguities may also disappear if one can observe blending shifts and identify the
blend (Alard et al.~\cite{Alard}) or
resolve the motion of the centroid of light due to the motion and brightening of the images
(H{\o}g et al.~\cite{HNP}; Paczy{\'n}ski~\cite{PacCentr}; Boden et al.~\cite{BSV}).  

This paper shows that it is of importance to look for all 
possible fits. The models shown (and by Dominik \& Hirshfeld (\cite{DoHi2}) for MACHO LMC\#1)
here give some insight to the possible configurations which arise for the
presented type of models.

Especially for claiming the existence of a planet from a microlensing light
curve, one has to be careful and study all possible models and consider the uncertainties
of the fit parameters as given e.g. by 1-$\sigma$-bounds (see also Gaudi \& Gould~\cite{GG}; 
Gaudi~\cite{Gaudi}). From these
fit parameters, information about the physical quantities (mass, separation)
can be obtained (Dominik~\cite{DoMass}).

\begin{acknowledgements}
I would like to thank S.~Mao for some discussions on the subject, 
C.~Alard for sending me the data of the DUO\#2 event, the OGLE collaboration
for making available their data, and A. C. Hirshfeld for reading the draft manuscript.
\end{acknowledgements}

\clearpage
\LARGE

\vspace*{5cm}

Figure 1
\vspace*{1cm}

\epsfig{file=H0433.f1}
\clearpage

\vspace*{2cm}

Figure 2a
\vspace*{1cm}

\epsfig{file=H0433.f2a}

\vspace*{2cm}

Figure 2b
\vspace*{1cm}

\epsfig{file=H0433.f2b}
\clearpage

\vspace*{2cm}

Figure 2c
\vspace*{1cm}

\epsfig{file=H0433.f2c}

\vspace*{2cm}

Figure 2d
\vspace*{1cm}

\epsfig{file=H0433.f2d}
\clearpage

\vspace*{2cm}

Figure 2e
\vspace*{1cm}

\epsfig{file=H0433.f2e}

\vspace*{2cm}

Figure 2f
\vspace*{1cm}

\epsfig{file=H0433.f2f}
\clearpage

\vspace*{2cm}

Figure 2g
\vspace*{1cm}

\epsfig{file=H0433.f2g}

\vspace*{2cm}

Figure 2h
\vspace*{1cm}

\epsfig{file=H0433.f2h}
\clearpage

\vspace*{5cm}

Figure 3a
\vspace*{1cm}

\epsfig{file=H0433.f3a}
\clearpage

\vspace*{5cm}

Figure 3b
\vspace*{1cm}

\epsfig{file=H0433.f3b}
\clearpage

\vspace*{5cm}

Figure 3c
\vspace*{1cm}

\epsfig{file=H0433.f3c}
\clearpage

\vspace*{5cm}

Figure 3d
\vspace*{1cm}

\epsfig{file=H0433.f3d}
\clearpage

\vspace*{5cm}

Figure 3e
\vspace*{1cm}

\epsfig{file=H0433.f3e}
\clearpage

\vspace*{5cm}

Figure 3f
\vspace*{1cm}

\epsfig{file=H0433.f3f}
\clearpage

\vspace*{5cm}

Figure 3g
\vspace*{1cm}

\epsfig{file=H0433.f3g}
\clearpage

\vspace*{5cm}

Figure 3h
\vspace*{1cm}

\epsfig{file=H0433.f3h}
\clearpage

\vspace*{5cm}

Figure 4a
\vspace*{1cm}

\epsfig{file=H0433.f4a}
\clearpage

\vspace*{5cm}

Figure 4b
\vspace*{1cm}

\epsfig{file=H0433.f4b}
\clearpage

\vspace*{5cm}

Figure 4c
\vspace*{1cm}

\epsfig{file=H0433.f4c}
\clearpage

\vspace*{5cm}

Figure 4d
\vspace*{1cm}

\epsfig{file=H0433.f4d}
\clearpage

\vspace*{5cm}

Figure 5a
\vspace*{1cm}

\epsfig{file=H0433.f5a}
\clearpage

\vspace*{5cm}

Figure 5b
\vspace*{1cm}

\epsfig{file=H0433.f5b}
\clearpage

\vspace*{5cm}

Figure 6
\vspace*{1cm}

\epsfig{file=H0433.f6}
\clearpage

\vspace*{1.5cm}

Figure 7a
\vspace*{1cm}

\epsfig{file=H0433.f7a}

\vspace*{1.5cm}

Figure 7b 
\vspace*{1cm}

\epsfig{file=H0433.f7b}
\clearpage

\vspace*{1.5cm}

Figure 7c
\vspace*{1cm}

\epsfig{file=H0433.f7c}

\vspace*{1.5cm}

Figure 7d
\vspace*{1cm}

\epsfig{file=H0433.f7d}
\clearpage

\vspace*{1.5cm}

Figure 7e
\vspace*{1cm}

\epsfig{file=H0433.f7e}

\vspace*{1.5cm}

Figure 7f
\vspace*{1cm}

\epsfig{file=H0433.f7f}


\begin{thebibliography}{}
\bibitem[1995]{Alard}
Alard C., Mao S., Guibert J., 1995, A\&A 300, L17
\bibitem[1998a]{Planet1}
Albrow M., Beaulieu J.-P., Birch P., et al. (The PLANET collaboration), 1998a, The 1995 pilot campaign of
PLANET: Searching for microlensing anomalies through precise, rapid, round-the-clock monitoring,
accepted for publication in ApJ 509
\bibitem[1998b]{Planet2}
Albrow M., Beaulieu J.-P., Caldwell J. A. R., et al. (The PLANET collaboration), 1998b, 
1997 PLANET monitoring of anonmalous events: First detection of limb-darkening via microlensing,
Proceedings of the 4th International workshop on gravitational microlensing surveys, eds. J. Kaplan
\& M. Moniez
\bibitem[1997a]{MACHOExt}
Alcock C., Allen W. H., Allsman R. A., et al. (The MACHO and GMAN collaborations), 1997a, ApJ 491, 436
\bibitem[1997b]{MACHOOGLE1}
Alcock C., Allsman R. A., Alves D., et al. (The MACHO collaboration), 1997, ApJ 479, 119
\bibitem[1994]{MACHOOGLE2}
Bennett D. P., Alcock C., Allsman R. A., et al. (The MACHO collaboration), 1994,
Recent developments in gravitational microlensing and the latest MACHO results: Microlensing
towards the galactic bulge, Proceedings of 5th Annual Astrophysics Conference in Maryland, ed. S. Holt,
preprint astro-ph/9411114 
\bibitem[1995]{MACHOOGLE3}
Bennett D. P., Alcock C., Allsman R. A., et al. (The MACHO collaboration), 1995,
The MACHO project dark matter search, Proceedings of the Astronomical Society of the Pacific Symposium on
Clusters, Lensing, and the Future of the Universe, preprint astro-ph/9510104
\bibitem[1996]{MACHOBin}
Bennett D. P., Alcock C., Allsman R. A., et al. (The MACHO collaboration), 1996,
Nucl. Phys. Proc. Suppl. 51B, 152
\bibitem[1998]{BSV}
Boden A. F., Shao M., Van Buren D., 1998, ApJ 502, 538
\bibitem[1997]{DisPer}
Di Stefano R., Perna R., 1997, ApJ 488, 55
\bibitem[1995]{DoAstro}
Dominik M., 1995, A\&AS 109, 597
\bibitem[1996]{DoDiss}
Dominik M., 1996, Galactic Microlensing beyond the Standard Model, PhD thesis, Universit\"at Dortmund
\bibitem[1998a]{DoMass}
Dominik M., 1998a, A\&A 329, 361 
\bibitem[1998b]{DoRob} 
Dominik M., 1998b, A\&A 330, 963
\bibitem[1994]{DoHi1}
Dominik M., Hirshfeld A. C., 1994, A\&A 289, L31
\bibitem[1996]{DoHi2}
Dominik M., Hirshfeld A. C., 1996, A\&A 313, 841
\bibitem[1997]{Gaudi}
Gaudi B. S., 1997, Planet microlensing perturbations: True planets or binary source?, 
preprint astro-ph/9706268
\bibitem[1997]{GG}
Gaudi B. S., Gould A., 1997, ApJ 486, 85
\bibitem[1995]{HNP}
H{\o}g E., Novikov I. D., Polnarev A. G., 1995, A\&A 294, 287
\bibitem[1995]{MaoDis}
Mao S., Di Stefano R., 1995, ApJ 440, 22
\bibitem[1991]{MP}
Mao S., Paczy{\'n}ski B., 1991, ApJ 374, L37
\bibitem[1997]{PacRot}
Paczy{\'n}ski B., 1997, Binary source parallactic effect in gravitational micro-lensing,
preprint astro-ph/9711007
\bibitem[1998]{PacCentr}
Paczy{\'n}ski B., 1998, ApJ 494, L23
\bibitem[1994]{Rhie}
Rhie S. H., 1994, contributed talk in the conference ``Sources of dark matter in the universe'',
16-18 February (unpublished)
\bibitem[1996]{RB}
Rhie S. H., Bennett D. P., 1996, Nucl. Phys. Proc. Suppl. 51B, 86 
\bibitem[1997]{MACHOAlert}
Stubbs C., et al. (MACHO collaboration), 1997, {\tt http://darkstar.astro.washington.edu}
\bibitem[1994]{OGLE7}
Udalski A., Szyma{\'n}ski M., Mao S., et al., 1994, ApJ 436, L103


\end{thebibliography}
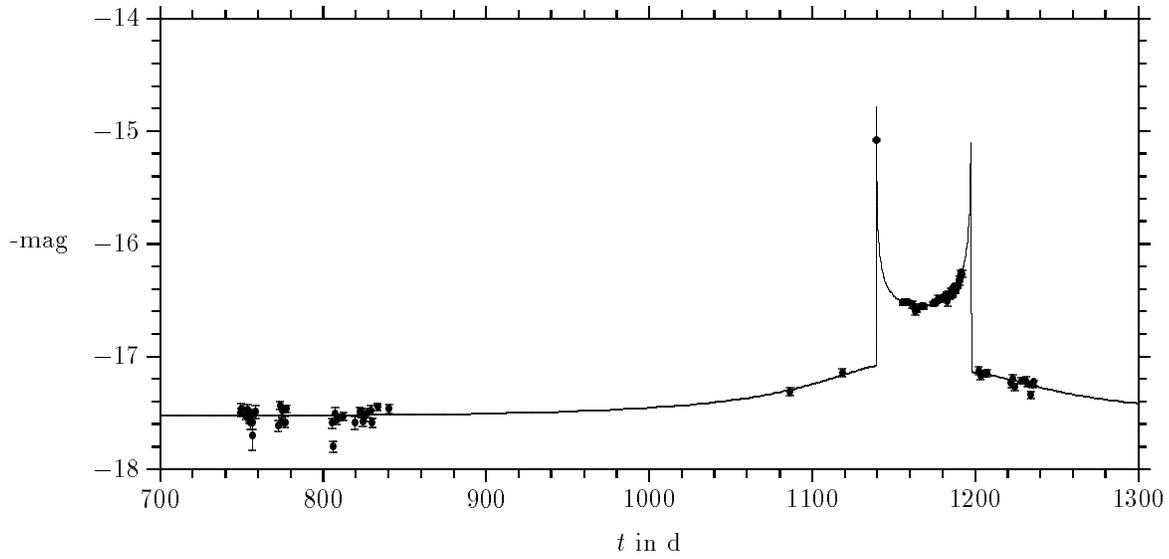
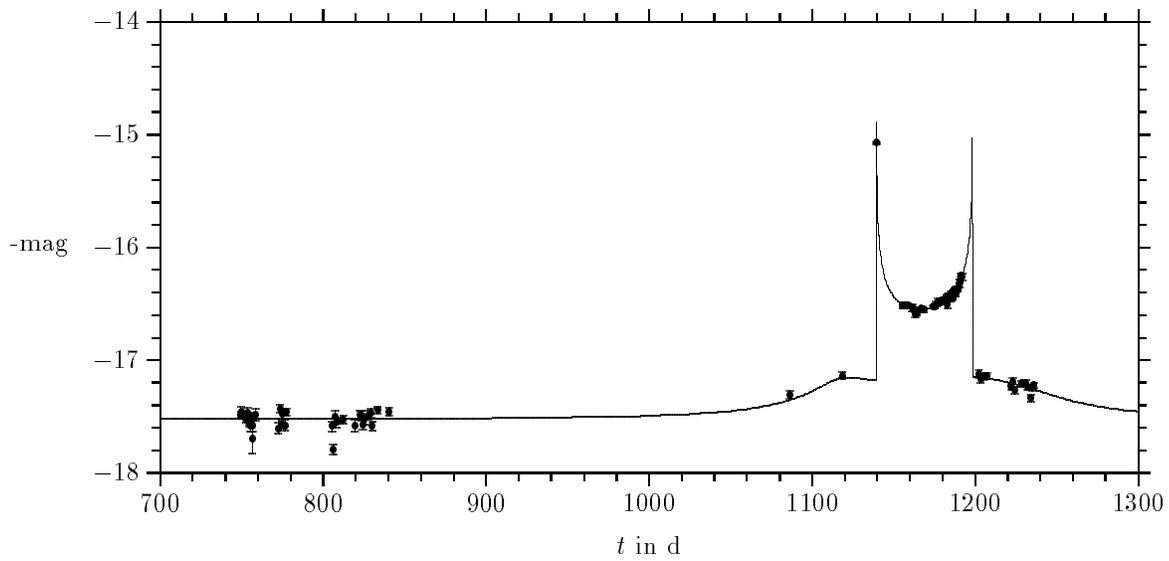
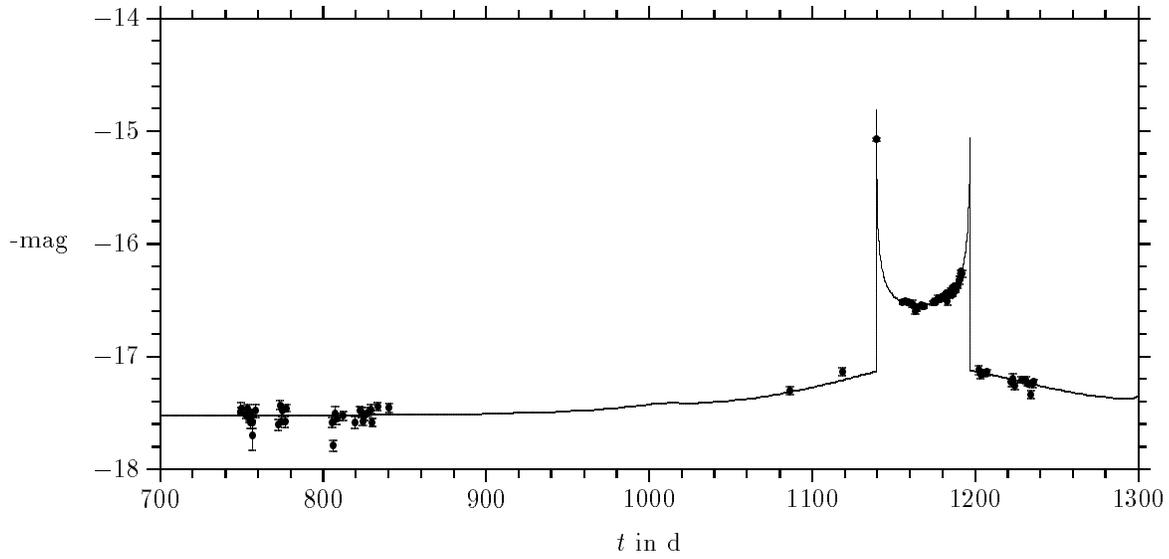
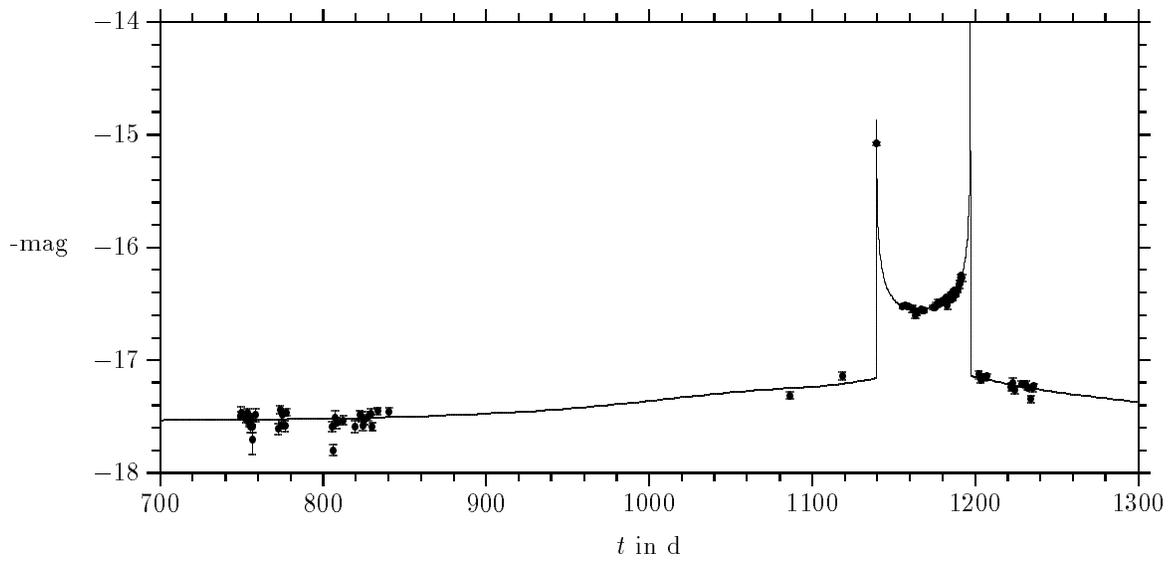
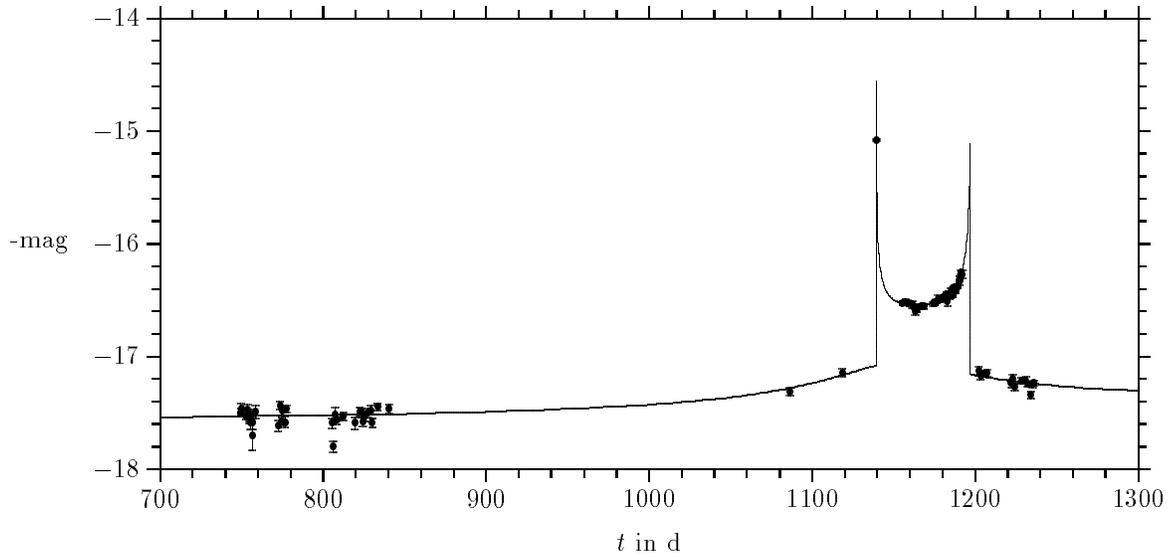
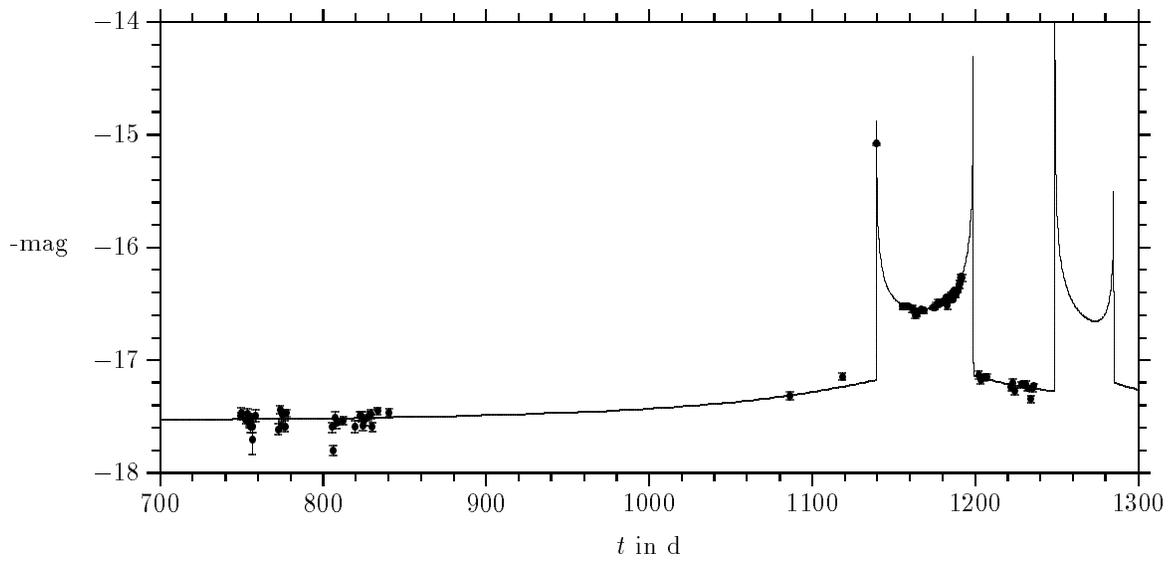
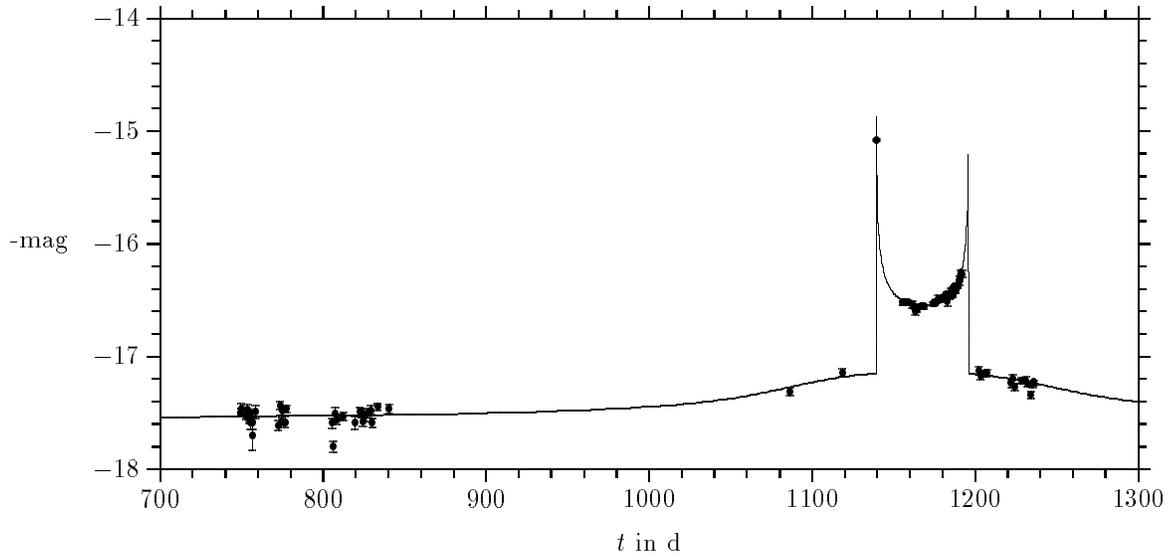
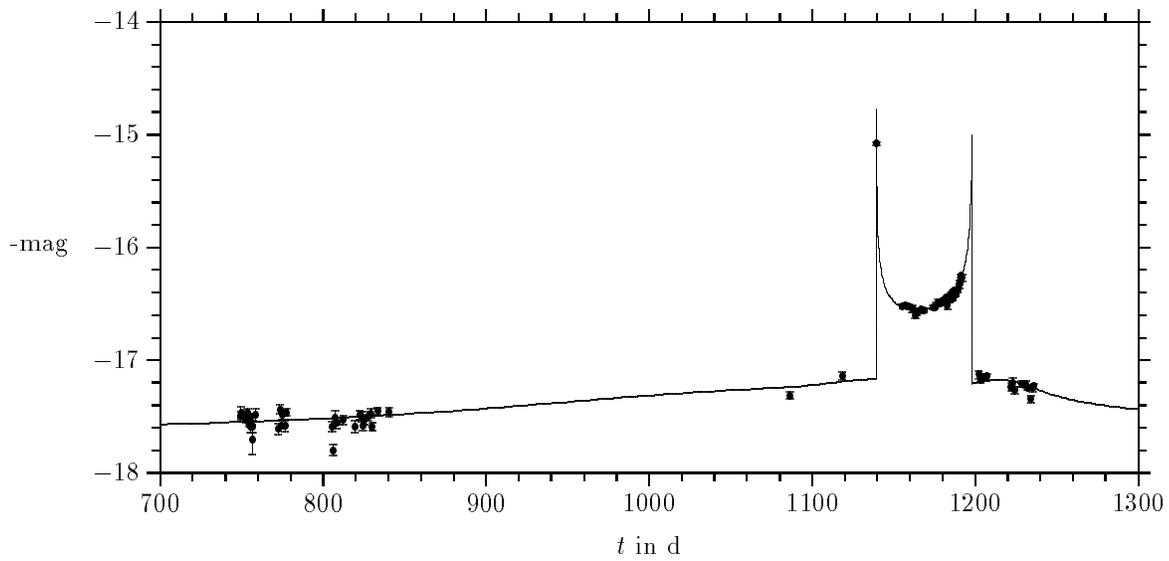
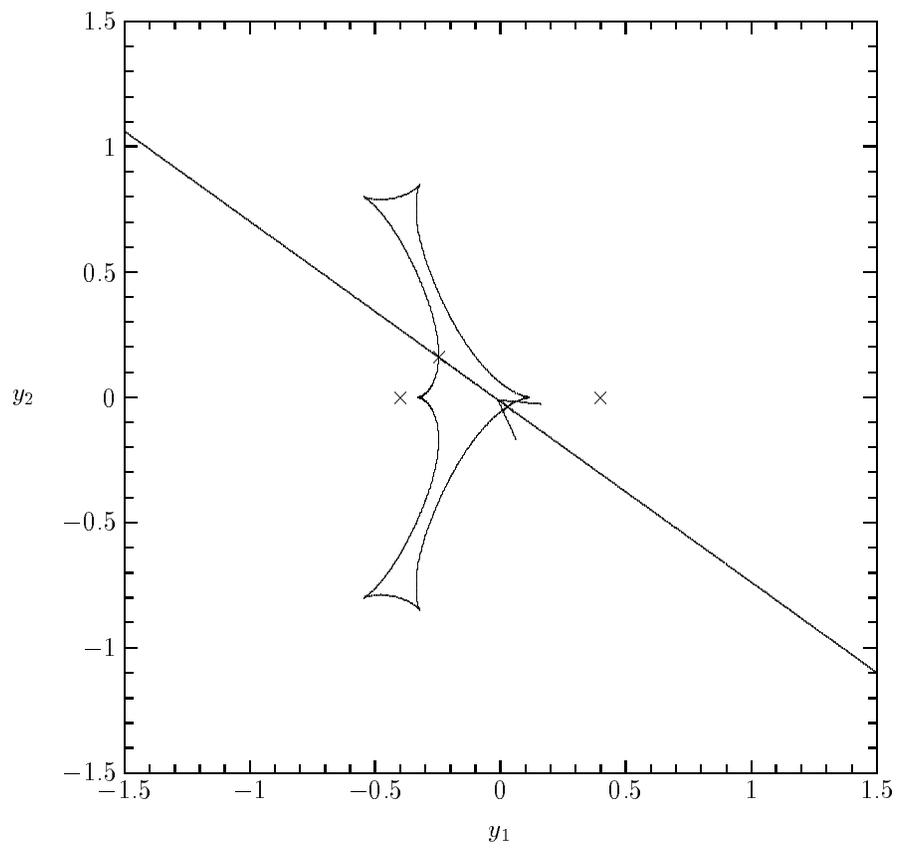
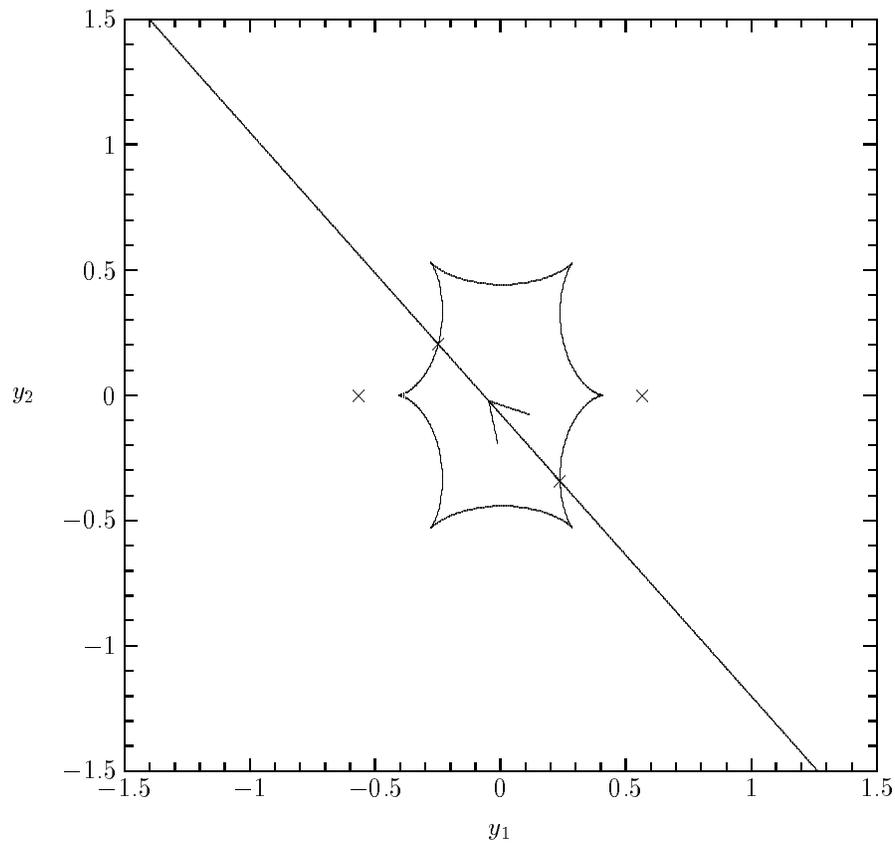
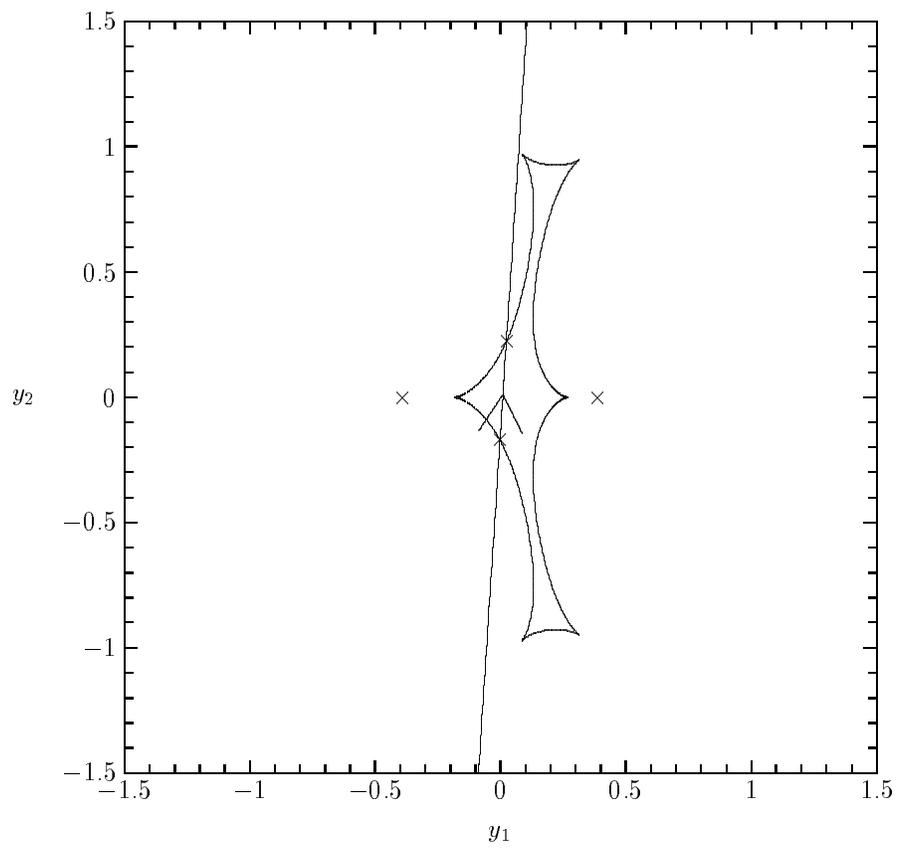
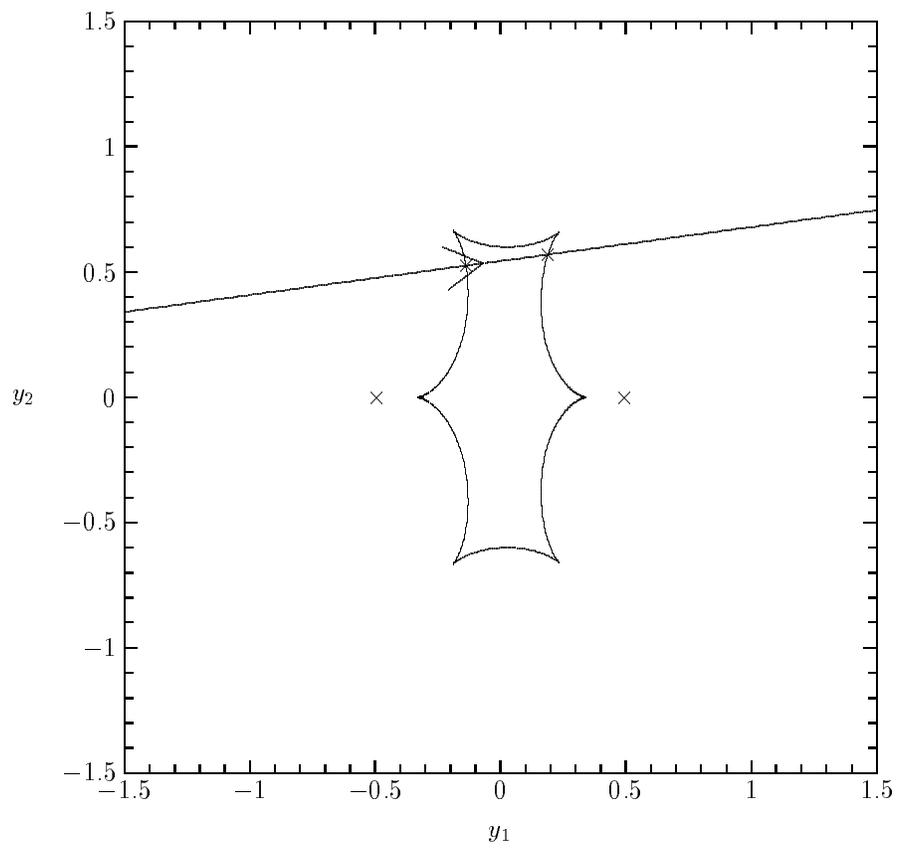
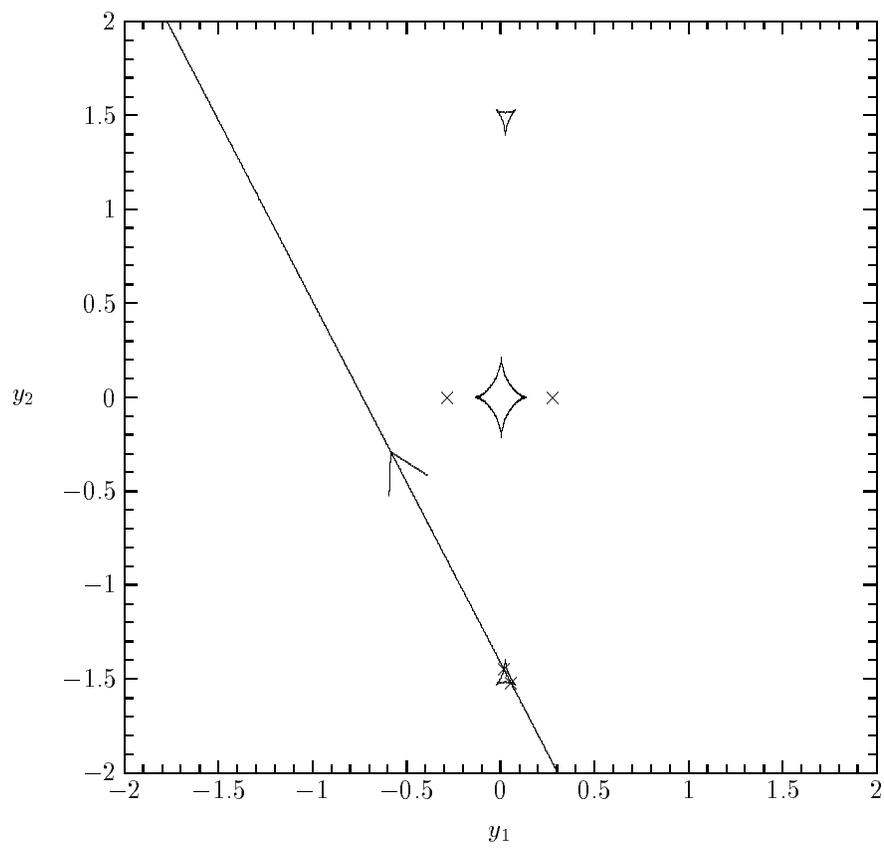
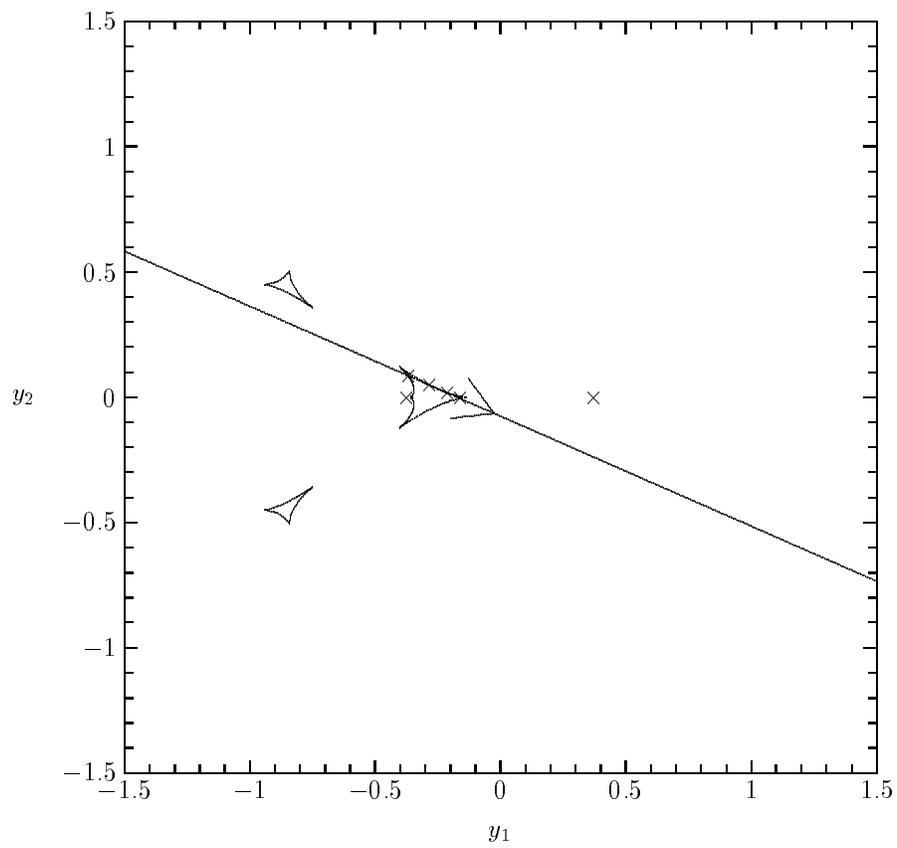
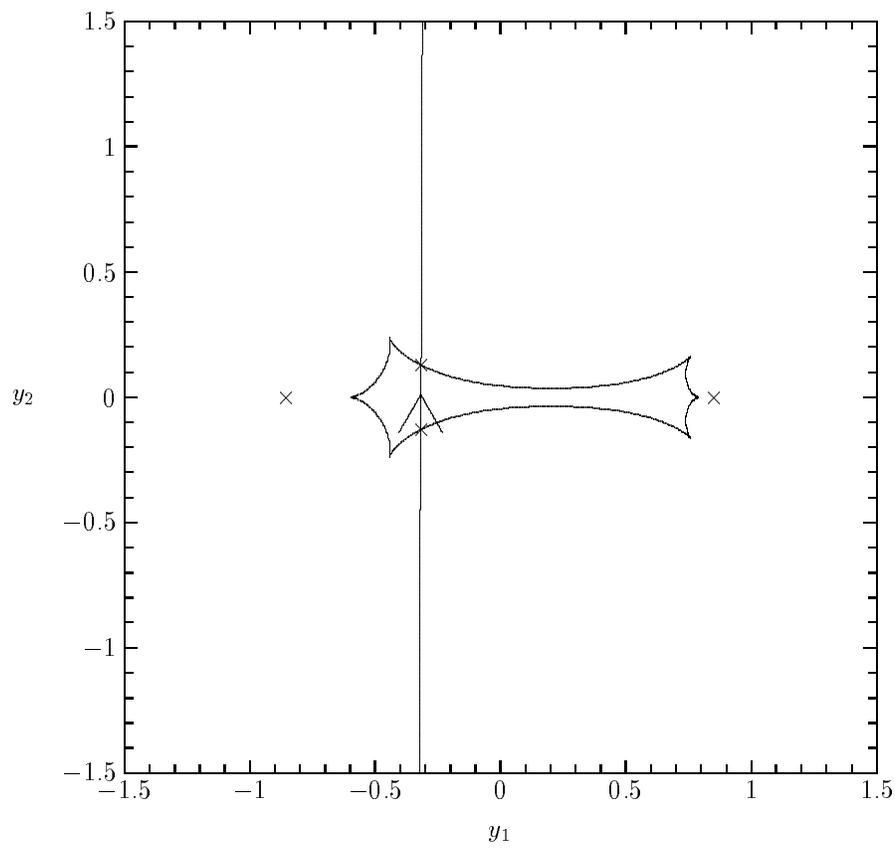
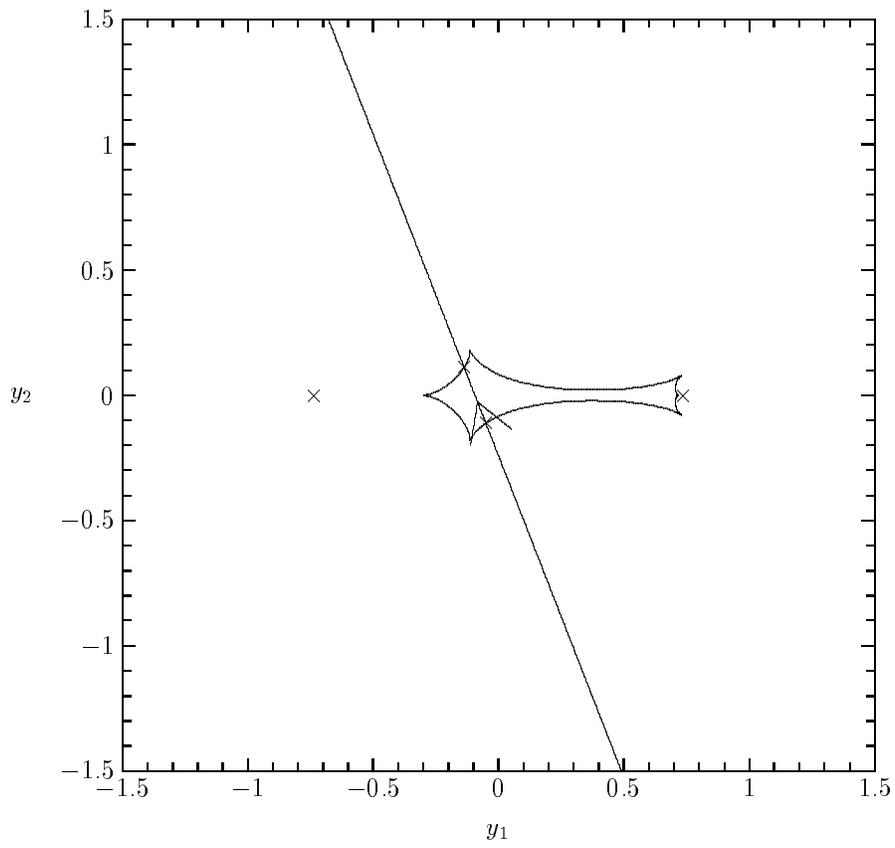
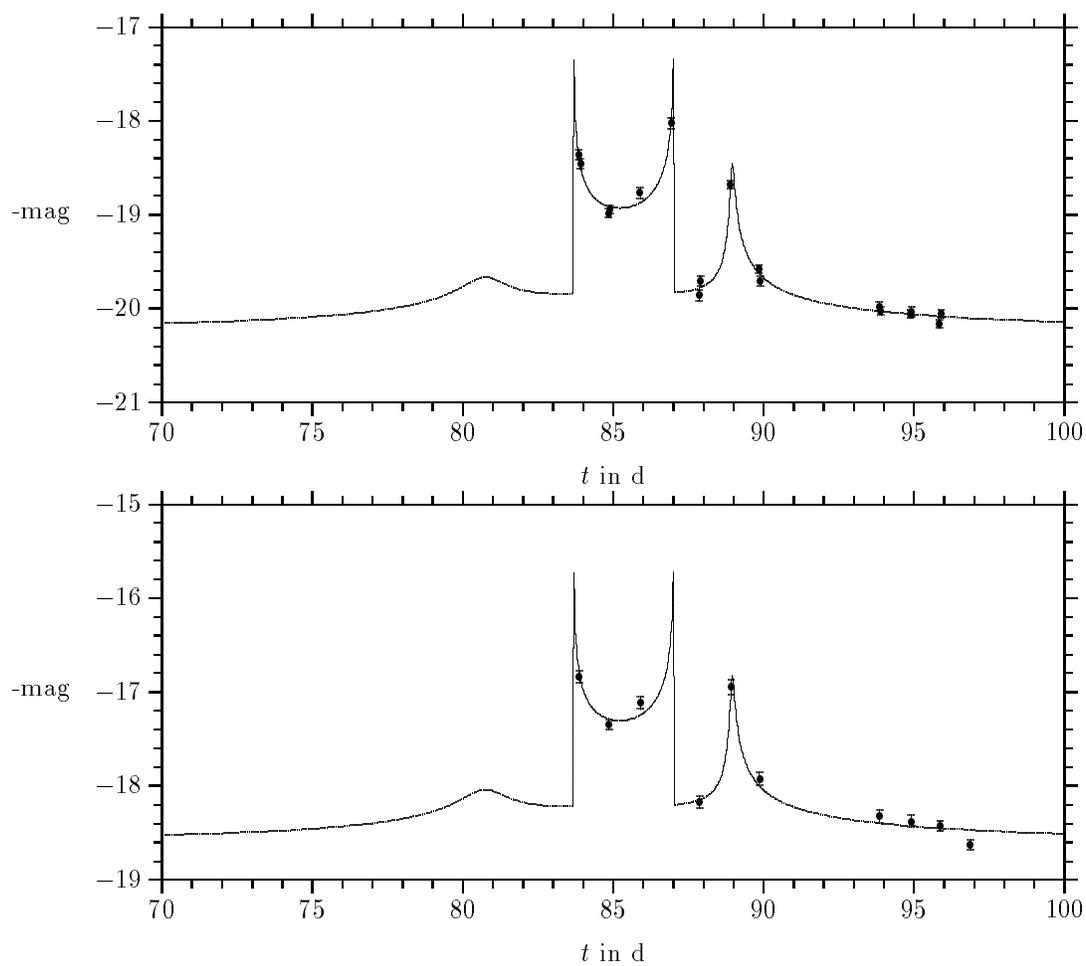
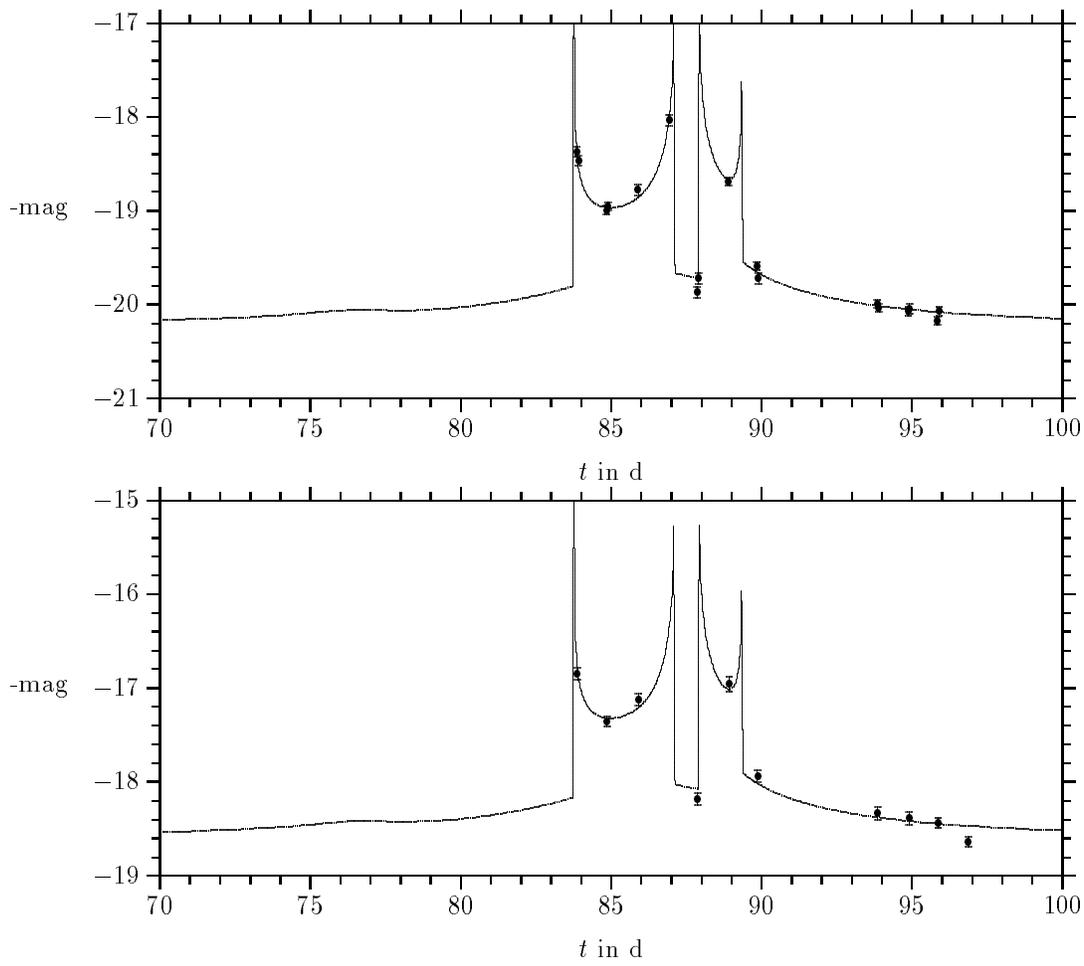
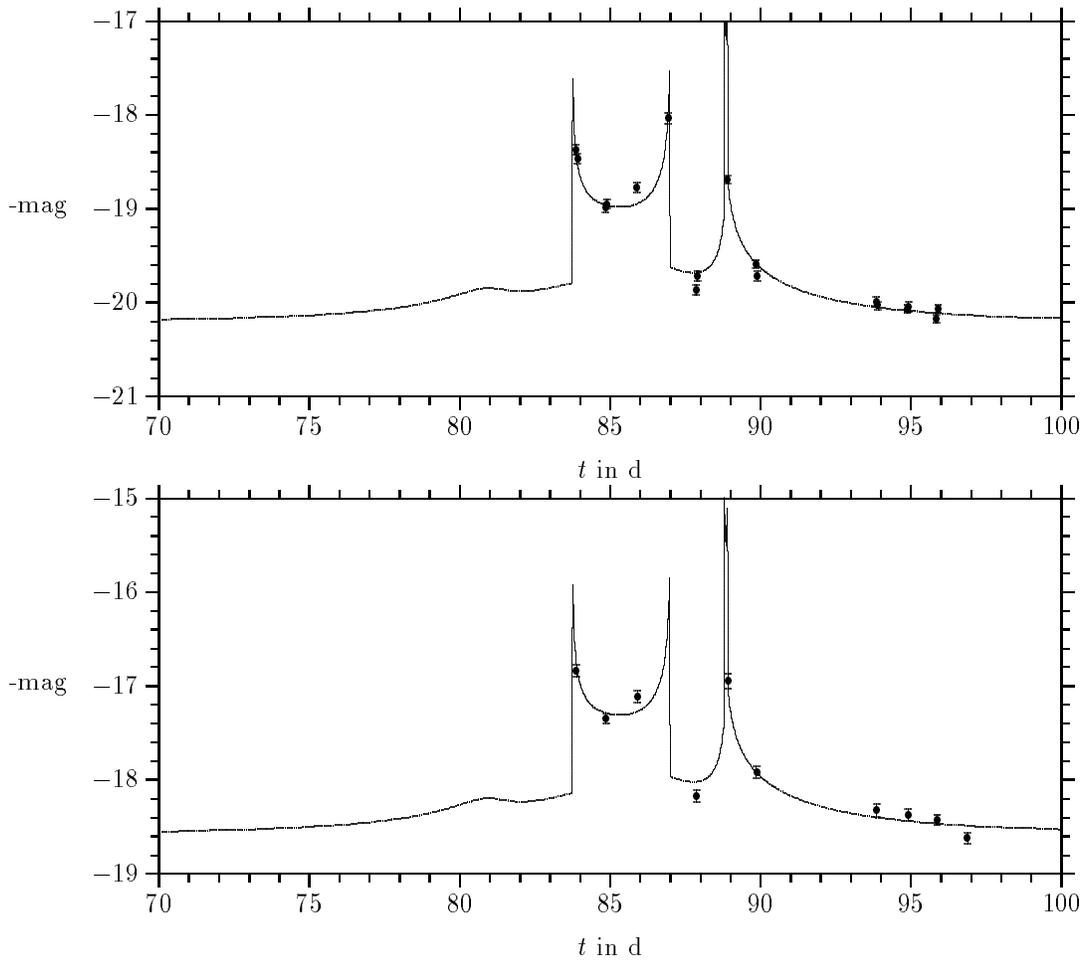
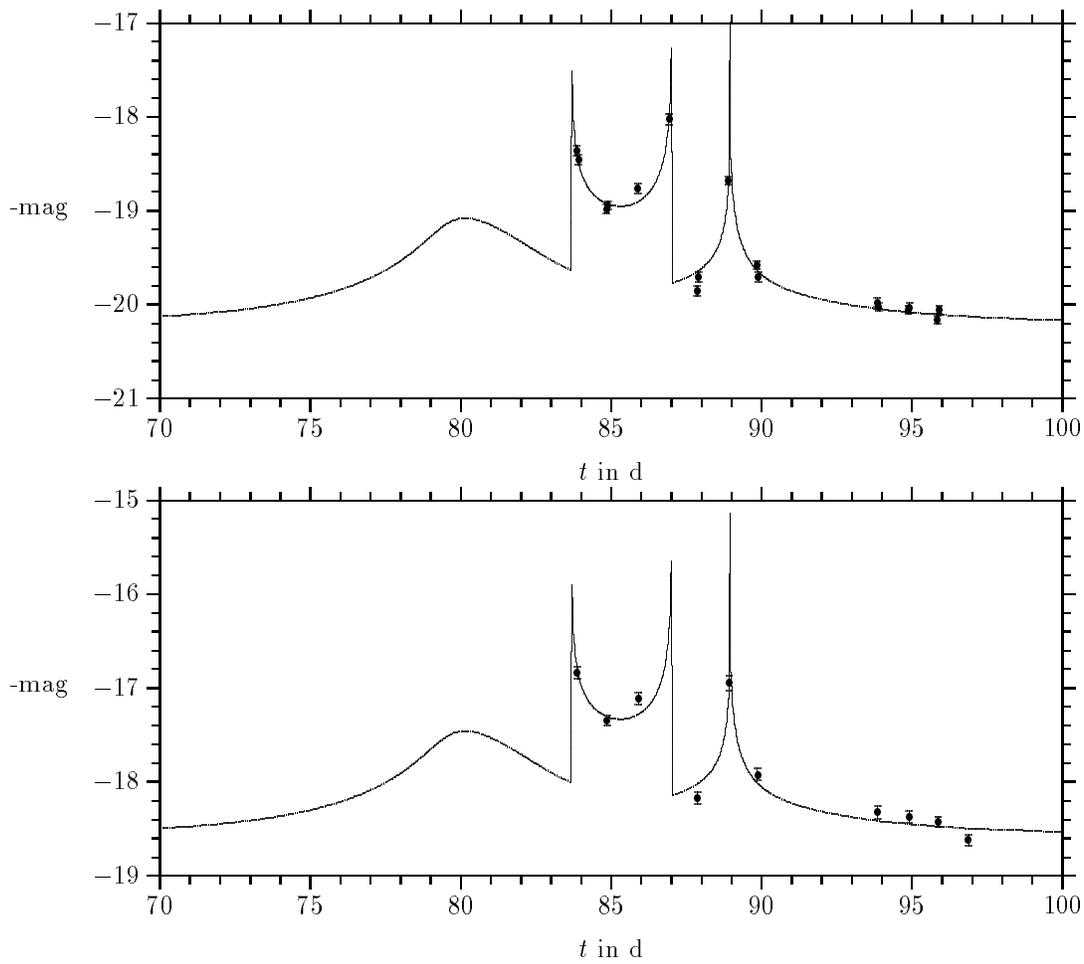
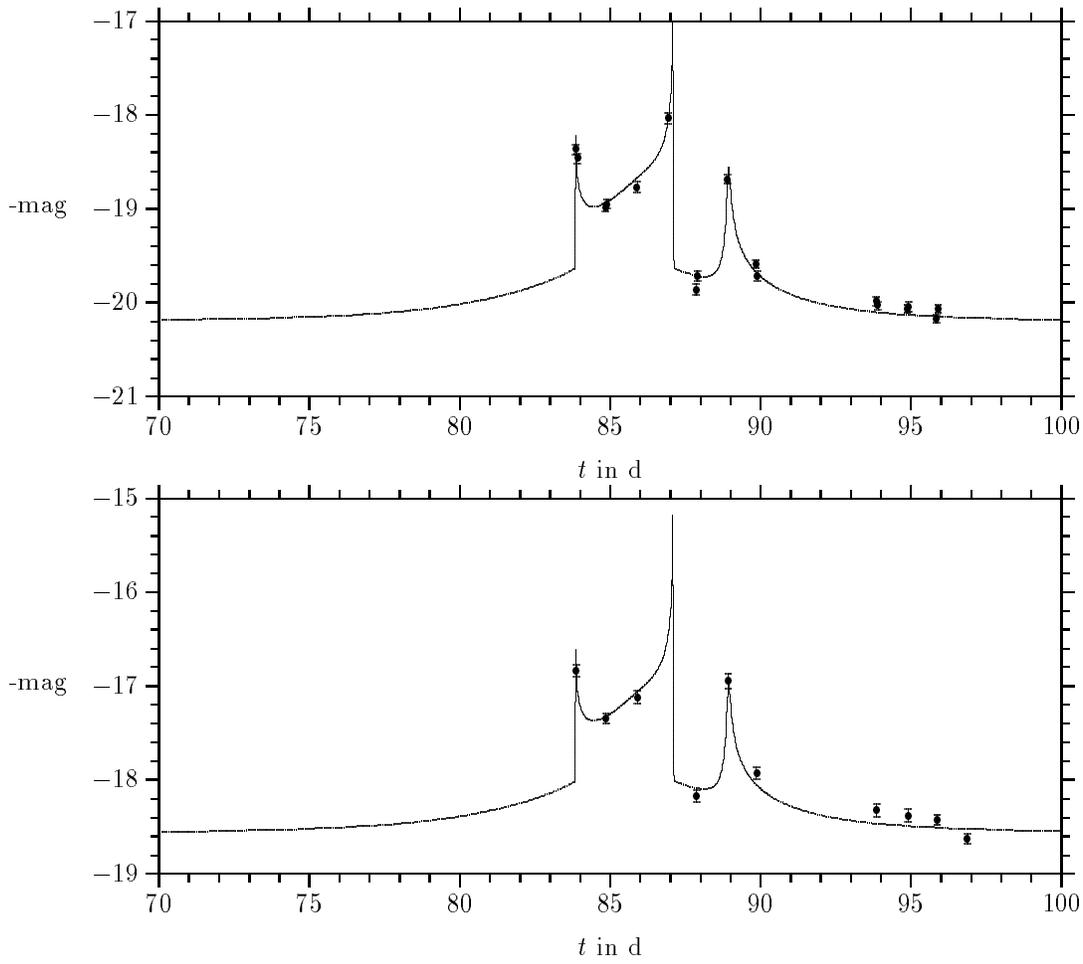
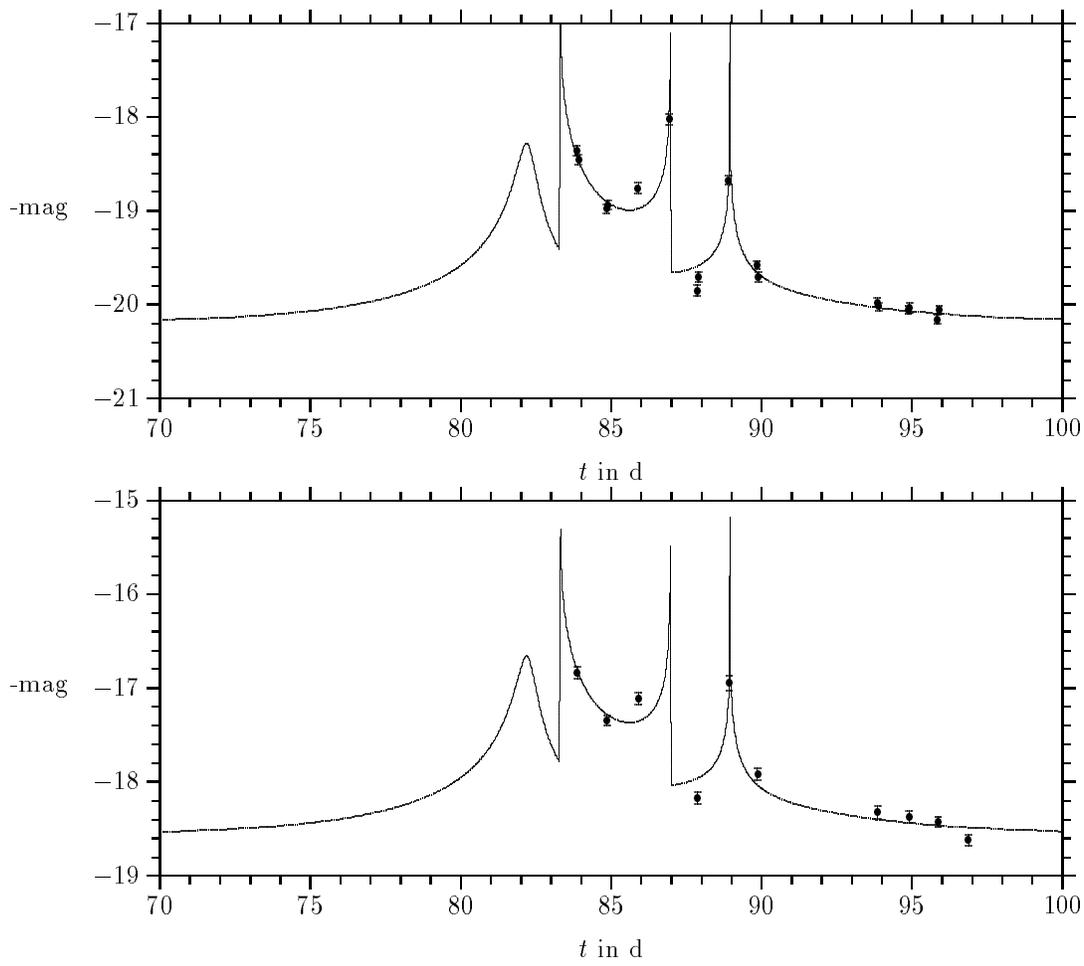
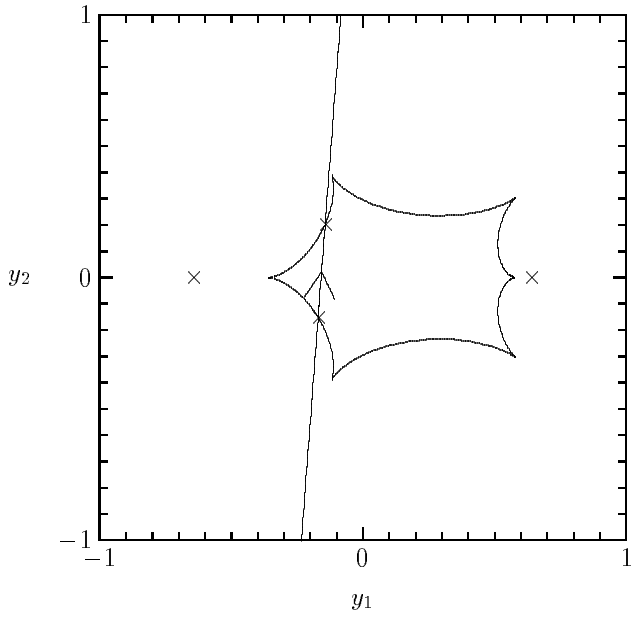
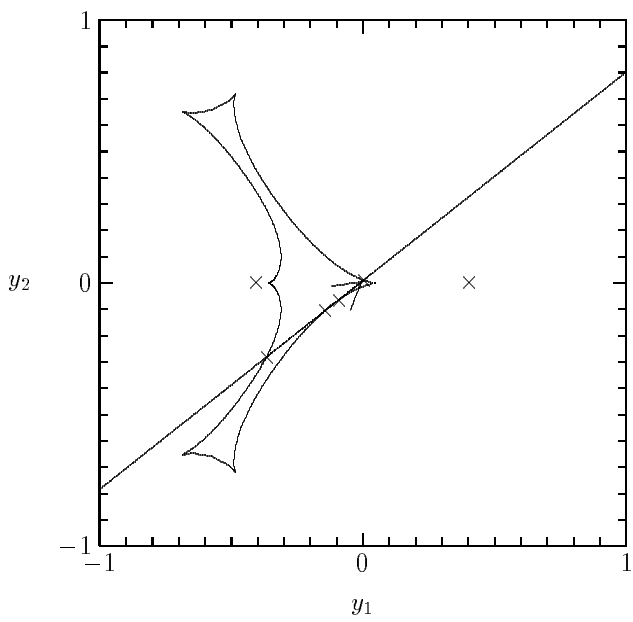
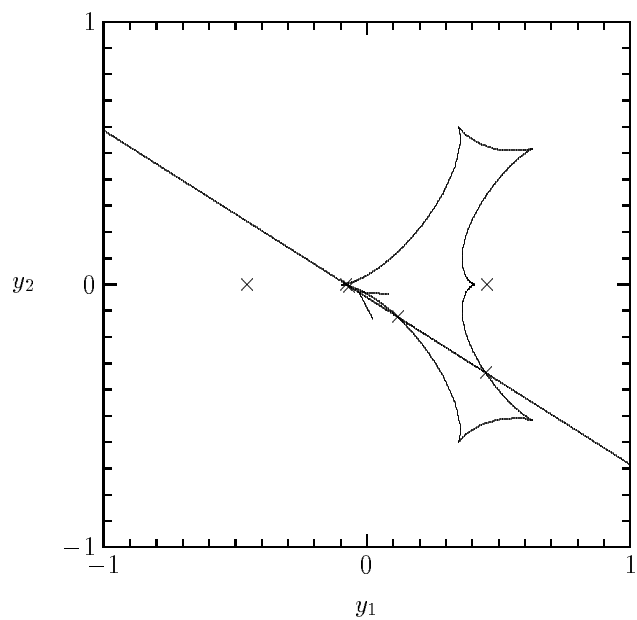
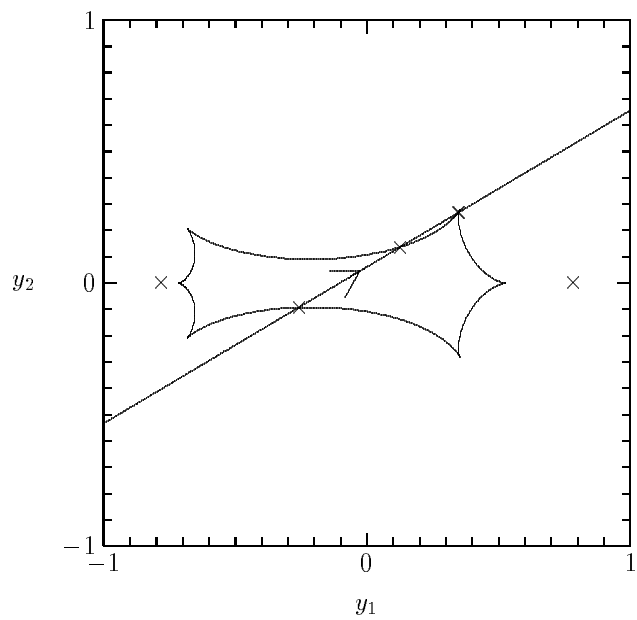
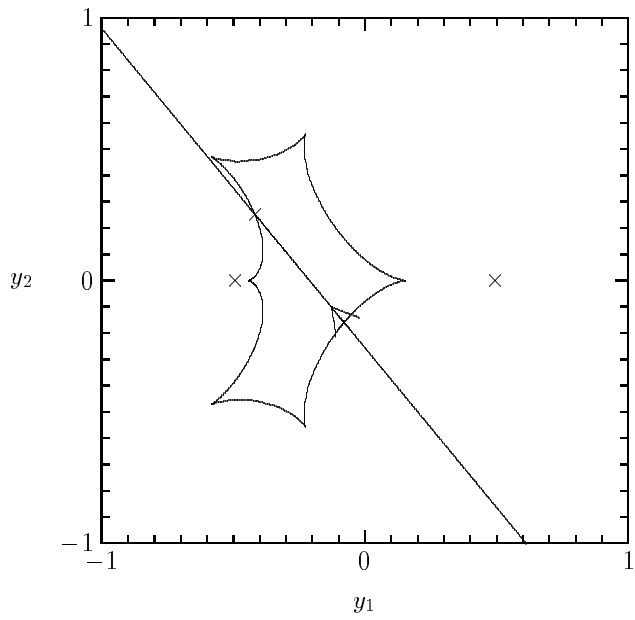
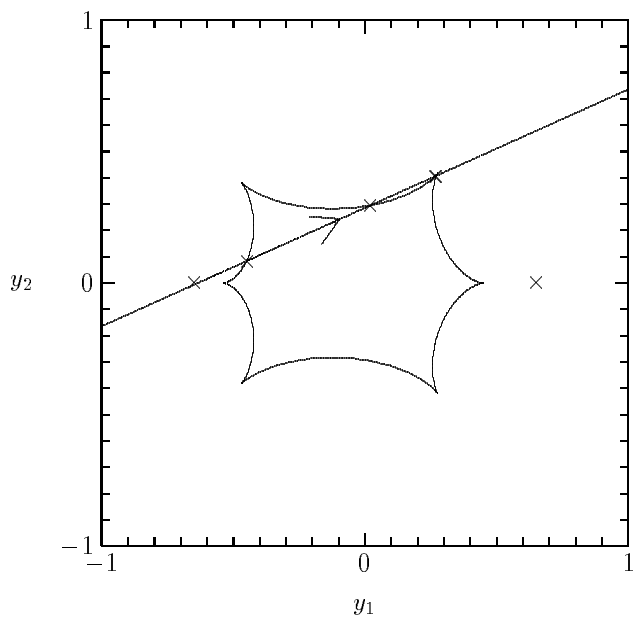
\end{document}